\documentclass[twocolumn]{svjour3}     %

\usepackage[sort&compress,numbers]{natbib}

\usepackage{graphicx}
\usepackage{amssymb}
\usepackage{amsmath} %
\usepackage{aas_macros} %
\usepackage{xcolor} %
\usepackage{soul} %
\usepackage{hyperref}
\hypersetup{
  unicode=false,         %
  pdftoolbar=true,        %
  pdfmenubar=true,        %
  pdffitwindow=true,       %
  pdfstartview={FitH},      %
  pdftitle={Nucleosynthesis and Observation of the Heaviest Elements}, %
  pdfauthor={eholmbeck},     %
  pdfsubject={Nuclear Astrophysics},  %
  pdfkeywords={nuclear, physics, actinides, r-process, stars, elements, superheavy, fission}, %
  pdfnewwindow=true,       %
  colorlinks=true,        %
  linkcolor=orange,       %
  citecolor=teal,         %
  filecolor=magenta,       %
  urlcolor=cyan,         %
  breaklinks=true,
  linktocpage
}

\begin{document}

\title{Nucleosynthesis and observation of the heaviest elements
}

\author{E.~M.~Holmbeck \and
    T.~M.~Sprouse \and
    M.~R.~Mumpower
}

\institute{
E.~M.~Holmbeck \at 
  The Observatories of the Carnegie Institution for Science, Pasadena, CA 91101, USA \\
  NHFP Hubble Fellow \\
  \email{eholmbeck@carnegiescience.edu}
  \and
T.~M.~Sprouse \at
Theoretical Division, Los Alamos National Laboratory, Los Alamos, NM 87545, USA \\
Center for Theoretical Astrophysics, Los Alamos National Laboratory, Los Alamos, NM, 87545, USA \\
       \email{tmsprouse@lanl.gov}      %
      \and
M.~R.~Mumpower \at
Theoretical Division, Los Alamos National Laboratory, Los Alamos, NM 87545, USA \\
Center for Theoretical Astrophysics, Los Alamos National Laboratory, Los Alamos, NM, 87545, USA \\
       \email{mumpower@lanl.gov}      %
}

\date{Received: 29 July 2022 / Accepted: 19 January 2023 / Published online: 15 February 2023}

\maketitle

\begin{abstract}
The rapid neutron capture or `\emph{r} process' of nucleosynthesis is believed to be responsible for the production of approximately half the natural abundance of heavy elements found on the periodic table above iron (with proton number $Z=26$) and all of the heavy elements above bismuth ($Z=83$).
In the course of creating the actinides and potentially superheavies, the \emph{r} process must necessarily synthesize superheavy nuclei (those with extreme proton numbers, neutron numbers or both) far from isotopes accessible in the laboratory.
Many questions about this process remain unanswered, such as `where in nature may this process occur?' and `what are the heaviest species created by this process?' 
In this review, we survey at a high level the nuclear properties relevant for the heaviest elements thought to be created in the \emph{r} process.
We provide a synopsis of the production and destruction mechanisms of these heavy species, in particular the actinides and superheavies, and discuss these heavy elements in relation to the astrophysical \emph{r} process.
We review the observational evidence of actinides found in the Solar system and in metal-poor stars and comment on the prospective of observing heavy-element production in explosive astrophysical events.
Finally, we discuss the possibility that future observations and laboratory experiments will provide new information in understanding the production of the heaviest elements.

\end{abstract}

\setcounter{tocdepth}{2}
\tableofcontents
\newpage

\section{Introduction}
\label{sec:introduction}

\begin{figure*}
 \centering
 \includegraphics[width=0.9\textwidth]{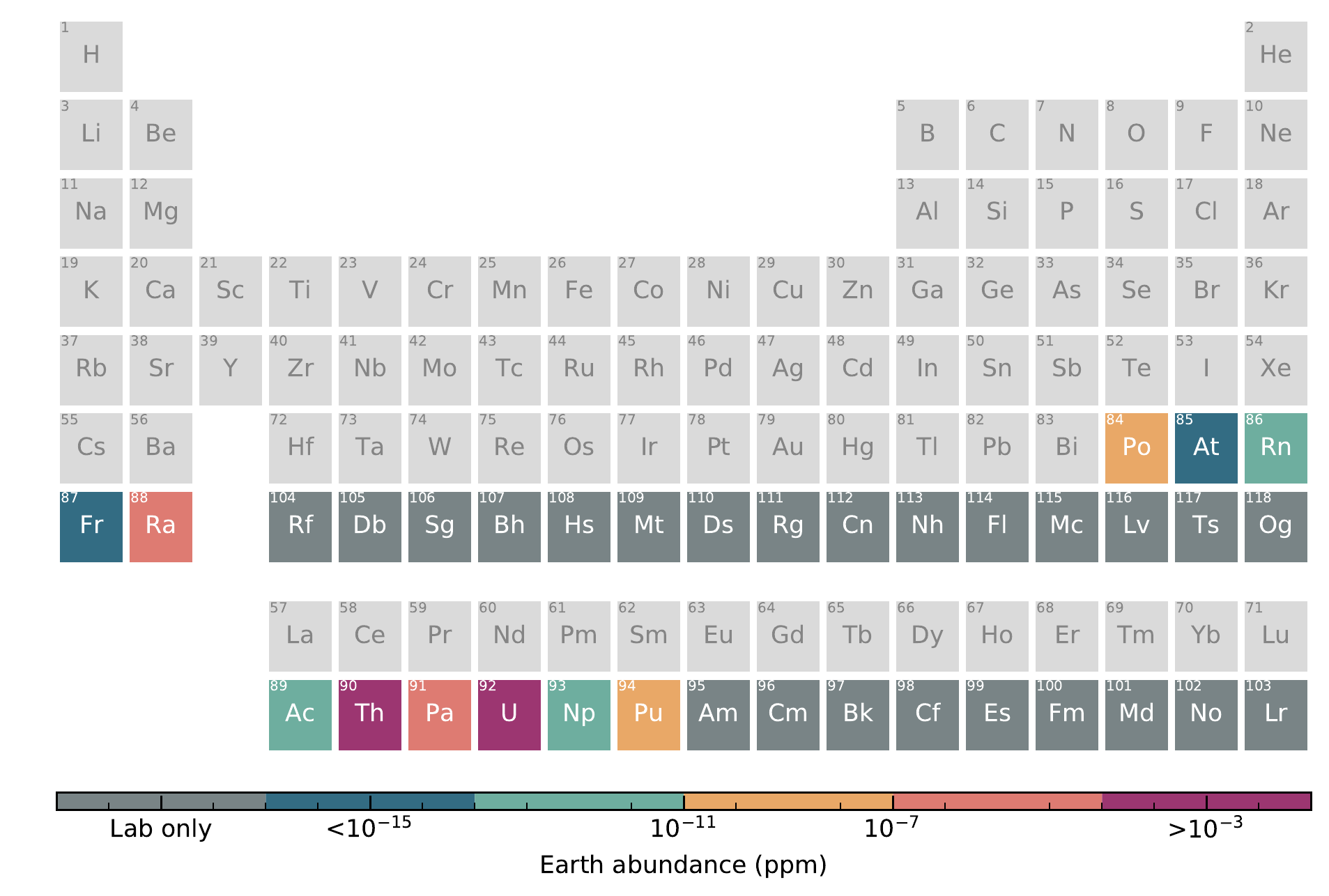}
\caption{Periodic table of the elements showing the Earth abundances of the heaviest elements above $Z>83$, including the superheavy elements ($Z>103$). Elements in dark grey are not found on Earth, but instead have been detected in laboratory settings.}
\label{fig:earth_heavies}
\end{figure*}

In the universe, we know of at least 118 chemical elements that can exist.
To visualize and sort these elements by their chemical properties, they are placed on the periodic table and grouped according to their chemical structure (see Fig.\ \ref{fig:earth_heavies}).
The heaviest elements on the periodic table are those with charge numbers in excess of $Z=83$, including the actinide series ($89\leq Z\leq 103$) and the transactinide---often called `superheavy'---elements with $Z>103$.
Experimental efforts over the past several decades have resulted in the discovery of new elements, including, at the present time, the heaviest known element oganesson, Og, with $Z=118$ \cite{Karol2016,Karol2016b}.
While all but two elements with $Z\leq 83$ have at least one state in which they live longer than a Hubble time ($\gtrsim$ 14 Gyr), all of the known elements above $Z=83$ are radioactive; that is, they only exist only for a finite amount of time.
However, an `Island of Stability' \cite{Myers1966,Nilsson1969,Moller1986} has been proposed for elements near $Z\sim 114$, where some of these elements are thought to be less radioactive and exist for longer than their superheavy neighbors.

Because heavy elements tend to decay quite quickly when compared to the time that has elapsed since their astrophysical production, most of the heavy elements---especially the transactinides---cannot be found naturally; instead, these may be created in laboratory settings.
There are a handful of actinide elements that live sufficiently long to be found naturally in significant amounts: thorium (Th), protactinium (Pa), uranium (U), neptunium (Np), and plutonium (Pu), with $Z=90$--94, respectively.
Figure \ref{fig:earth_heavies} highlights the actinide and transactinide elements on the periodic table and colors them by their abundance on Earth.
In addition, all of the elements between Bi and Th ($83< Z< 90$) occur naturally in the Earth's crust as decay products of the longer-lived actinides, though only in trace amounts.
Beyond Earth, Th and U are the only $Z>83$ elements that can also be identified in the light of quiescent stars beyond our Solar system, notably in some of the oldest stars in the Galaxy.

The presence of the heavy elements both inside and outside the Solar system attest to a natural production mechanism for these elements \cite{Kajino2019}.
However, no \emph{trans}-actinide element has yet been found in nature, neither in the Solar system nor outside of it.
Can they be created naturally at all? 
Can the process that makes the actinides also create the superheavies, and is there any chance of finding or observing these elements in their natal production sites? 
To address these questions, this review first introduces in Sec.\ \ref{sec:nucphys} the nuclear physics involved in the production (and destruction) of the heaviest elements.
Next, Sec.\ \ref{sec:nucleosynthesis} reviews where and how these nuclear properties are active in the astrophysical context that creates the heavy elements.
We follow up with the observational consequences of heavy-element nucleosynthesis in Sec.\ \ref{sec:observation}, discussing the possibilities of observing the heaviest elements and their natural production.
Lastly, Sec.\ \ref{sec:summary} offers encouraging perspectives as we move into the future of this topic at the intersection of physics and astronomy.

\subsection{From atoms to nuclei}

We have, up until this point in our review, focused on the periodic table as the taxonomy within which the rich variety of atomic nuclei may be assigned.
Nuclei may be grouped according to their overall number of protons into specific elements.
With relatively few exceptions, atomic nuclei belonging to the same element will share a common electronic structure and, consequently, share a common set of chemical and material properties, such that their shared proton number presents an excellent basis for their classification throughout much of chemistry, material science, engineering, and even our daily interactions with the world.

Beginning around 100 years ago, it became increasingly apparent that the classification of atoms according to their chemical properties \textit{alone} was insufficient for uniquely identifying atoms.
The gold foil experiments by Ernest Rutherford, Hans Geiger, and Ernest Marsden between 1908 and 1913 set the initial groundwork by developing an atomic model comprised of a dense, positively charged core (what we now know to be the nucleus) surrounded by a relatively dispersed volume occupied by oppositely charged electrons \cite{Rutherford1911,Geiger1913}.
Subsequent experiments over the following few decades---in particular, the discovery of the neutron---eventually developed the understanding that the atomic core was comprised of neutrons in addition to protons, giving rise to a further distinction between atoms of the same element \cite{Rutherford1913}.

The many years that followed these pioneering efforts in nuclear physics ultimately culminated in the %
modern understanding of the atomic nucleus: the atomic nucleus is a bound system of neutrons and protons, each of which are composite systems of elementary particles known as quarks.
These constituent quarks are bound together by the strong force and mediated by gluons.
The protons and neutrons themselves are similarly bound by the strong force but mediated by the exchange of mesons, a quark-antiquark composite particle.
The total number of protons ($Z$) in a nucleus is still broadly responsible for setting its \textit{chemical} properties, but the widely variable number of neutrons ($N$) that may be combined with those protons to form bound nuclei of various \emph{isotopes} calls for an independent axis along which they may be sorted, namely neutron number.
Similarly, a nucleus can be referred to by its `atomic mass number,' $A$: the total combined number of neutrons and protons in the nucleus.
In this scheme, a \textit{nuclide} refers to a unique combination of some number of protons and neutrons in a bound system and is the nuclear analogue to the chemical `element.' 
As the elements are arranged on the periodic table according to their increasing proton number and electron shell structure, nuclides are similarly arranged on a \textit{chart of nuclides}, with axes not only describing the number of protons in the nucleus, but also the number of neutrons, indicating all the isotopes that may exist of a particular element.

We define superheavy \textit{nuclei} to refer to those nuclei which possess an extreme number of protons, neutrons or combination thereof; broadly categorizing nuclei with $Z > 82$.
Superheavy \textit{elements} refer to those species with $Z > 103$, forming a subset of superheavy nuclei.
The definition of a superheavy element depends on the atomic or nuclear context; here we take the nuclear perspective.

An element's proton number and electron shell structure (visualized by its location on the periodic table) are the basic properties that describe how an element reacts chemically.
Similarly, in addition to its proton and neutron number identity, there are a variety of properties of the nucleus that determine how individual isotopes react and decay.
These range from the relatively straightforward, such as their mass or binding energy, to the more complex, such as the precise description of how a nucleus may interact with other nuclei.
While any attempt to enumerate these properties in their entirety requires a devoted, exhaustive review, we dedicate Sec.\ \ref{sec:nucphys} to a more focused discussion of the small subset of these properties which are especially important for understanding the astrophysical production of the heaviest elements.

\subsection{Nuclear transmutations and nucleosynthesis}
\label{sec:intro:nucleosynthesis}

Insofar as all nuclei are comprised of the same basic building blocks, protons and neutrons, it immediately follows that there may be some prospect that nuclei may be able to \textit{re}configure themselves into new combinations of the same inherent building blocks.
Indeed, this general phenomenon has been observed since the infancy of nuclear physics---the observation of radioactivity being perhaps the earliest example---even if it would not be understood as a transmutation of nuclei \textit{per se} for many years later.

Nuclear transmutations may be categorized in any number of ways.
One of the more natural ways to approach this task is to consider the circumstances under which a transmutation takes place.
For example, some nuclei transmute \textit{spontaneously}, in the sense that they proceed on their own without interaction with any particles external to the nucleus itself.
Spontaneous fission and $\alpha$ decay are clear examples of such transmutations.
These spontaneous decays differ from \textit{induced} transmutations, which only occur when a nucleus interacts with some external particle, such as a baryon, another nucleus, or perhaps a lepton.

Another commonly used basis for describing nuclear transmutations concerns the types of interactions that mediate the transmutation.
As it is ultimately the strong interaction that binds nuclei together, a broad swath of nuclear transmutations are principally driven in some complex way by the same category of interactions.
To highlight how this might proceed, consider the situation in which an incident (`projectile') nucleus bombards another (`target') nucleus.
In this scenario, it is possible for some protons and/or neutrons be exchanged between the two.
After such a transmutation takes place, the overall numbers of protons and neutrons are independently conserved, but they are nevertheless rearranged to convert the initial target and projectile nuclei into some other nuclei.
It is not, however, the case that \textit{all} nuclear transmutations conserve the numbers of protons and neutrons, as weak interactions may convert protons to neutrons and vice-versa, the prototypical example of which is that of nuclear $\beta^{\pm}$ decay, which changes the element of a nucleus but leaves the combined number of neutrons and protons unchanged.
Similarly, not all reactions that preserve proton and neutron number are necessarily mediated by the strong force, e.g., photodistintegration is mediated by the electromagnetic force.

While the distinction between strong and weak transmutations is useful for discussing such topics at a high level, the fundamental forces are always active, and thus the distinction is not always clean-cut.
For example, nuclear $\beta$ decay (a weak process) leaves a nucleus in an excited state which may emit a proton or neutron (via properties of the strong interaction) before finding its way to its final energy state.
Another example is $\alpha$ decay which exhibits an interplay between strong and electromagnetic interactions.
In general, a thorough understanding of all transmutations in which a nucleus might participate will require detailed consideration of all types of interactions as a collective whole.

Given the rich and diverse types of transmutations in which nuclei may participate, the question is raised: do these transmutations actually occur in any significant way in physical environments? 
This is perhaps most evident when we consider that the immediate aftermath of the Big Bang is understood to have produced primordial nuclei comprised almost entirely of hydrogen, helium, and lithium, in stark contrast with what we now observe to exist today, both in the Solar system and throughout the universe (as shown, e.g., in Fig.\ \ref{fig:earth_heavies}).
In contrast, early studies of Big Bang nucleosynthesis predicted the \emph{entire} range of elements were created \cite{Alpher1948}.
Given the modern understanding that the Big Bang only produced the lightest elements, the logical explanation for the observation of the rest of the elements found presently would be that nuclear transmutations have been taking place throughout the universe since its very beginning, converting the light nuclei into successively heavier ones, up to and including the naturally occurring thorium, uranium, and plutonium observed, e.g., here on Earth.
`Nucleosynthesis' is the generic term that refers to any process which produces new nuclei in physical systems, with `astrophysical nucleosynthesis' referring specifically to nucleosynthesis that takes place in astrophysical environments.

\begin{figure}
 \centering
 \includegraphics[width=\columnwidth]{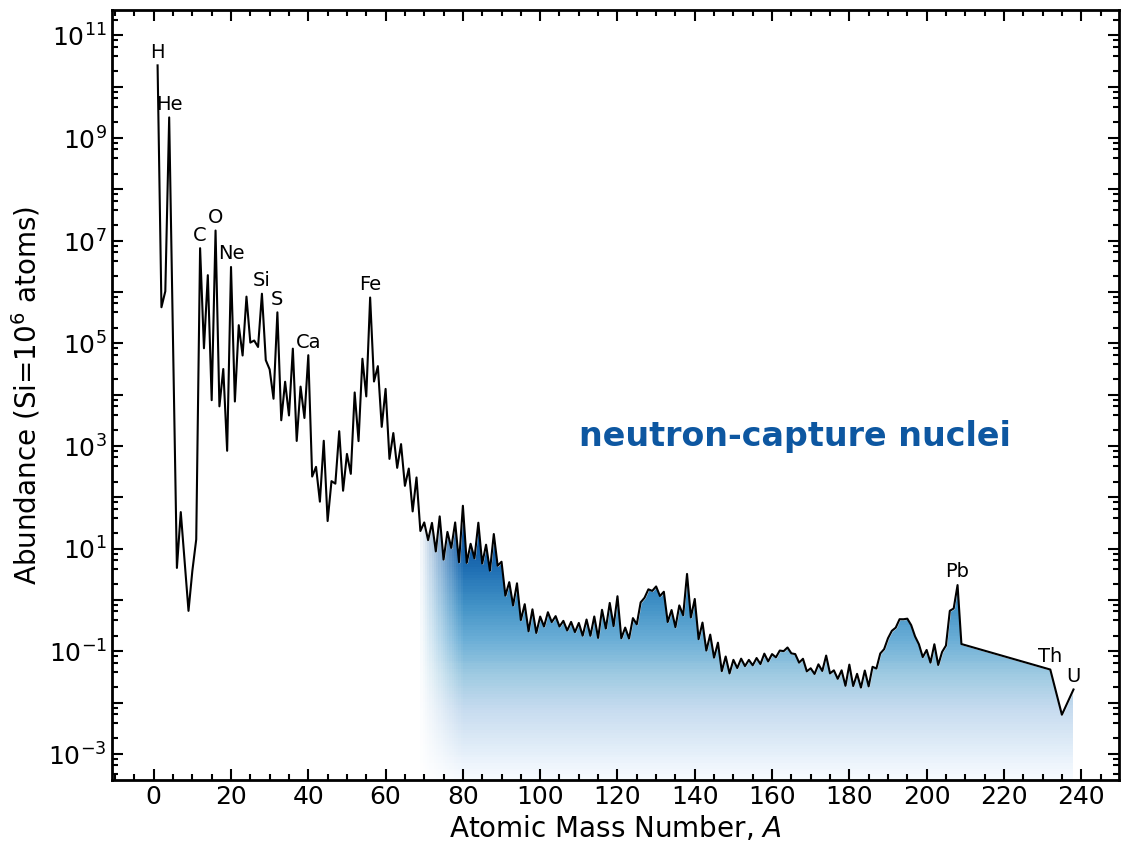}
 \caption{Abundance of each isobar (constant proton number plus neutron number, $A$) in the Solar system from data in Ref.\ \cite{Lodders2009}. The nuclei with primary origins in neutron-capture nucleosynthesis are highlighted in blue.}
 \label{fig:generic_solar}
\end{figure}

The elements we find in our Solar system, then, are a product of roughly nine billion years of distinct and incredibly diverse astrophysical nucleosynthesis processes that created the heavier elements out of the building blocks left by the Big Bang. Figure \ref{fig:generic_solar} shows the abundance of each isobar (those nuclei sharing the same total number of protons and neutrons in their nucleus, $A$) found in the Solar system on a logarithmic scale.
While hydrogen and helium are still the most abundant elements---reflective of their primordial origin---the rest of the periodic table ($Z\geq 3$) accounts for only about a total of 2\% of the Solar system material \cite{Lodders2003,Lodders2009}.

As can be seen in Fig.\ \ref{fig:generic_solar}, there is a steady decline in abundance with increasing atomic mass number, where the heaviest elements are also the least abundant.
A surprisingly robust theory for explaining this pattern as a consequence of nucleosynthesis was first proposed in the landmark 1957 result of Burbidge, Burbidge, Fowler, and Hoyle \cite{Burbidge1957} and contemporaneous works, e.g., by Al Cameron \cite{Cameron1957}.
These works, which gave birth to the field now known as `nuclear astrophysics,' interpreted the distinct peaks in the Solar system abundance pattern as a clear signature of nuclear physics structure and stellar evolution.
The peaks of locally high abundances, decreasing through the series of C, O, Ne, Si, and S, correspond to the $\alpha$-nuclei preferentially formed in the fusion stages of progressively more massive stars, terminating in the iron peak around $A=56$ produced by core-collapse supernovae that mark the fusion end-point of very massive stars' lives.
Beyond this point, i.e., moving towards the heavier nuclei populating the Solar system (indicated by the shaded region of Fig.\ \ref{fig:generic_solar}), the Coulomb barrier increases for heavier elements, making charged particle reactions increasingly unsustainable in stellar environments.
In other words, the fusion of like-charged nuclei (e.g., Fe and Fe), which for elements lighter than Fe produced energy that once supported the star, now become energetically expensive for elements heavier than Fe, leading to stellar collapse.
It follows, then, that the elements beyond Fe cannot be produced astrophysically by stellar \emph{fusion}.
Rather, the capture of free neutrons onto nuclei is responsible for the vast majority of the abundances of elements with $Z\gtrsim 30$ in the Solar system, though charged-particle captures are still responsible for the production of some Solar-system nuclei.
However, because free neutrons decay with a half-life of around ten minutes, the nucleosynthesis of these heaviest nuclei requires specific and relatively rare astrophysical environments: that is, an environment which also produces the neutrons to be subsequently captured.
The capture of light charged nuclei (e.g., a proton) is a possible route to synthesizing heavier neutron-deficient elements.
The rapid capture of protons (\emph{r}p process) \cite{Wallace1981} is thought to produce light \emph{p} nuclei up to mass number $A\sim110$, see \cite{Schatz2001} and references therein.
The astrophysical conditions (temperature and density) must be appropriate to favor incident protons overcoming the Coulomb barrier of heavier nuclei.
In environments with sufficient neutrino fluxes, anti-neutrino capture on protons, or the $\nu$p process, may ensue.
This process can explain the production of $^{92,94}$Mo and $^{96,98}$Ru in the Solar system \cite{Frolich2006}.
As far as the Solar system is concerned, the heavy elements owe the majority of their production to two generic types of neutron capture: slow and rapid \cite{Meyer1994}.
Any prospect for the production of the (super)heavy elements would necessarily proceed via \emph{rapid} neutron-capture nucleosynthesis (or the `\emph{r} process'), a concept which we explore in much deeper detail in Sec.\ \ref{sec:nucleosynthesis}.

\subsection{Hints of heavy element nucleosynthesis}

In parallel to breakthroughs in the understanding of atoms and nuclei were landmark identifications of heavy elements on Earth.
Today we know that trace amounts of several heavy elements produced by the \emph{r} process can be found on Earth (see Fig.\ \ref{fig:earth_heavies}).
Uranium was first discovered in 1789 from mineral samples \cite{Klaproth1789} and $^{244}$Pu more recently in deep-sea sediments \cite{Hoffman1971,Wallner2015,Wallner2021}.
(For an exhaustive review of the history of actinide discoveries, see \cite{Fry2013} and \cite{Kragh2013}.) 
Even before the direct detections of the actinide elements, the existence of Th, U, and Pu was inferred from measurements of Xe and Pb as the radioactive `mother' nuclei that have undergone nuclear transmuation by decaying into lighter `daughter' species (e.g., \cite{Chackett1950,Dalton1953,Kuroda1960}).
As early as the 1910's, going one nuclear generation further by requiring the origin of uranium as itself a decay daughter product of a still-heavier element was both hypothesized and criticized \cite{Haas1912,Kragh2013}.
As we will see in this review, this hypothesis was ironically not completely incorrect.
The actinides can indeed be decay products of the superheavy elements, and observations of the actinides may be an indirect route to inferring production of the superheavies \cite{Oganessian2006}.

The definition of `observation' is not limited to direct sample collection and measurement such as that conducted on Earth through, e.g., meteorites or deep-sea sediments; most of the usable and available information about the universe comes from starlight.
A star is essentially a cool layer of material surrounding a hotter core that radiates light over a continuum of energies (i.e., wavelengths).
Elements in the stellar photosphere can `absorb' photons radiated by deeper layers of the star at certain energies.
Specifically, if a photon has enough energy to excite an electron to a higher atomic energy level (or ionization state), the photon may be `absorbed' by the electron.
This exchange of energy manifests in the electromagnetic spectrum of the relatively cool stellar photosphere as an atomic `absorption feature' in which the brightness of specific wavelengths of light is less than its neighboring wavelengths that are not absorbed (or not absorbed as strongly).
In a quiescent star---such as the Sun---the hot stellar core produces a continuum of light at different wavelengths, and elements in the cooler atmosphere of the star may absorb light emitted by the hot core, producing absorption features unique to the elemental composition of the stellar photosphere.
Following this principle, many elements can be identified in the light from the Sun.
Indeed, helium was first discovered in 1868 from sunlight during a solar eclipse and thus owes its name to the Greek word for `sun,' \emph{helios} \cite{Haig1868,Rayet1868}.
Somewhat more recently---and more than one hundred year after its terrestrial discovery in 1828---thorium was first identified in the Solar spectrum \cite{Moore1943}, and we emphasize that this identification would not have been possible without dedicated `laboratory astrophysics' efforts to characterize the atomic transitions of thorium in order to guide where they can be found in starlight \cite{McNally1942}.
Outside the Solar system, the energy spectrum of starlight can also reveal the elements present in other stars.
In Sec.\ \ref{sec:observation}, we will discuss stellar observations of the heaviest elements in more detail.

Observations both on Earth and from the Sun's spectrum of light are able to capture the elemental composition of the Solar system to produce diagrams like Fig.\ \ref{fig:generic_solar}, showing how much of each element/isobar exists in the Solar system \cite{Arnould2020}.
Observations from light and those from physical samples are fundamentally different, however.
Atomic absorption features visible in stellar spectra are set by the proton number in the nucleus, and only in rare cases do the absorption features show sensitivity to the total number of nucleons in the nucleus.
In other words, abundances derived from starlight only indicate the total elemental abundance (i.e., summed over all isotopes that may be present) \cite{Roederer2017,Hansen2018}.
On the other hand, laboratory techniques are able to separate elements not only by their charge, but also by their mass number, providing another axis along which to characterize the elemental abundances and allowing abundances as a function of atomic mass number to be made (such as in Fig.\ \ref{fig:generic_solar}).
Because the Solar system is the only system for which isotopic measurements can be made across the whole observable periodic table, it remains the only system for which we have an \emph{isotopic} abundance pattern \cite{Cowan2021}.
Stardust from other systems arriving to the Solar system in the form of pre-Solar grains also allows some isotopic information to be gained, but these grains are uncorrelated with stars that are presently observed with stellar spectroscopy.
The abundance patterns of quiescent stars other than our Sun are strictly by element, save for the rare handful of elements for which isotopes can be identified in stellar spectra \cite{Wenyuan2018,Roederer2022}.

Similarly, atomic features (absorption or emission) can be detected in non-quiescent events as well.
For example, William Huggins made the first spectroscopic observations of a novae in 1866 \cite{Huggins1866}, noting that the hydrogen lines are notably different from those found in stars \cite{Hoffleit1986}.
Today, classes of novae and supernovae are categorized based on what elements are detectable in their spectra.
In Sec.\ \ref{sec:observation}, we will address if, when, and how evidence for the heaviest elements can be found in spectra of explosive astrophysical events.

\section{Nuclear physics of the heaviest nuclei}
\label{sec:nucphys}

The production of the heaviest elements in nature critically depends on the properties of very short-lived, unstable nuclei.
To date, many of these properties remain unmeasured, and therefore theoretical model calculations of nuclear properties must be used to supplement nuclear data when simulating nucleosynthesis taking place in extreme regions of the chart of nuclides.
In this section, we review the nuclear properties that are of particular interest to the formation of the heaviest elements.

\subsection{Nuclear binding energy and mass}\label{sec:masses}

One of the basic properties of a nucleus is its binding energy, or the minimal energy required to separate the nucleus into its constituent nucleons (neutrons and protons).
From the Einstein energy-mass relation, this can be equivalently stated in terms of mass.
In principle this quantity can be derived from the lowest energy state of the many-body Hamiltonian; however, this calculation is a tremendous task for light nuclei \cite{Carlson1998} and such complexities only increase for heavy species \cite{Koszorus2021,Mougeot2021}.

The first estimates of nuclear binding energy were based on the liquid drop model of a nucleus, and it was shown that this energy saturates per nucleon \cite{Gamow1930}.
Within a few years after this idea, a phenomenological formula was developed by Weizs\"{a}cker~\cite{Weizsacker1935}.
This semi-empirical formula consists of additive terms, each describing a unique aspect of the physics of nuclear binding energies:
\begin{equation}
\begin{split}
  E_{B}(Z,N) = a_{V}A - a_{S}A^{2/3} - a_{C}\frac{Z^2}{A^{1/3}}
             \\ - a_{A}\frac{(N-Z)^2}{A} + \delta(Z,N) \ ,
\end{split} 
\label{eqn:weizsaecker}
\end{equation}
where the first term describes the increase in binding energy as more nucleons are added to the nucleus, the second term describes the surface contribution of nucleons near the edge of the nucleus, the third (Coulomb) term describes the electrostatic repulsion of protons, the fourth term measures the asymmetry (Pauli exclusion), and the final term controls pairwise spin-coupling.
The coefficients ($a_i$) for each of these terms are fit to the measured nuclear masses of nuclei, such as those compiled and summarized in the Atomic Mass Evaluation (AME) \cite{Audi2003,Wang2017,Wang2021}.
Note that going from measured \emph{atomic} masses and \emph{nuclear} masses requires an additional term to account for the binding energy of the electrons.
The latest (2020) release of the AME dataset contained the atomic masses of roughly 2,500 species.
For the three latest releases of the AME, the average reported uncertainty among all nuclei was 40, 30, and 25 keV respectively.
From this progression it is clear that the precision of measurements has continually improved over time, particularly with the replacement of $\beta$ endpoint measurements with measurements from Penning traps \cite{Block2016}.

Perhaps the most immediately apparent shortcoming of the Weizs\"{a}cker mass formula is its inability to reproduce the so-called `closed shell' features appearing in nuclear masses.
These closed shells correspond to nucleons completely occupying discrete energy levels of the nucleus and are a direct parallel to the closed valence electron shells of the noble gas elements (i.e., the last column of Fig.\ \ref{fig:earth_heavies}).
By observing the nuclear masses as a function of proton and neutron number, one notes the relatively stronger binding energies of nuclei with specific numbers of neutrons and/or protons compared to their immediate neighbors on the chart of nuclides.
The canonical closed shells occur at nucleon `magic' numbers of 2, 8, 20, 28, 50, 82, and 126.
For heavy or deformed nuclei, the situation becomes somewhat more complicated, with the closed shells appearing at different numbers of protons and neutrons.
For example, the heaviest confirmed closed shell for the proton occurs at $Z=82$, with the next accessible closed shell suggested at 114; for the neutron, the heaviest closed shell is at $N=126$, and it is suggested that $N=184$ may also be a closed shell \cite{Sobiczewski1966,Myers1966}.
The shell behavior of heavy, neutron rich nuclei at $N=126$ and $N=184$ has a uniquely important role in mediating the production of the heaviest elements in astrophysical environments, which we discuss in more detail in Sec.\ \ref{sec:nucleosynthesis}.
Nuclei with both closed neutron and proton shells tend to be spherical and are considered `doubly magic.' 
The `Island of Stability'---a region of the nuclear chart in which superheavy elements are theorized to be long-lived---is conjectured to occur around the postulated doubly magic closed shell at $Z=114$ and $N=184$, though there is debate on where exactly on the $N$-$Z$ plane the Island of Stability would exist \cite{Mosel1969,Lalazissis1996,Cwiok1996,Gambhir2005,Santhosh2017b,Adri2021}.

While Eqn.\ \ref{eqn:weizsaecker} can reproduce basic patterns arising in the masses of measured nuclei, the growing expansion of our knowledge of the nuclear landscape demands the need for more complex theoretical descriptions of the atomic nucleus, both to capture the aforementioned nuclear shell effects and other, more subtle features of nuclear structure.
Below, we survey some of the more common approaches that have been applied to the heaviest nuclei in recent years.

The semi-classical approach---sometimes called the macroscopic-microscopic method---breaks the description of nuclear binding down into two parts: (1) the energies related to the macroscopic shape, a more refined counterpart to the Weizs\"{a}cker formula, and (2) the energies related to quantum phenomena.
The motivation for this approach is empirical in nature and stems from the mean-field picture of atomic nuclei \cite{Negele1982}.
Popular semi-classical models include the widely used Finite-Range Droplet Model (FRDM) \cite{Moller1995,Moller2012,Moller2016}, Dulflo-Zuker (DZ) \cite{Duflo1995}, KTUY \cite{Koura2005}, and Weizs\"{a}cker-Skyrme (WS) \cite{Wang2010,Liu2011,Wang2014}.
Much of the modern day understanding of neutron-rich nuclei relevant to the \emph{r} process comes from the fundamental insights provided by semi-classical models \cite{Nix1969,Sobiczewski1969,Moller1974,Krappe1979,Moller1981,Nix1989,Ichikawa2019}.
To date, macroscopic-microscopic approaches provide a match to mass evaluations on the order of a few hundred keV \cite{Mumpower2016r}.

A number of powerful and predictive methods seek to model the atomic nucleus based on the nucleon-nucleon interactions between its constituent protons and neutrons, e.g., no-core shell models \cite{Navratil2000,Navratil2016}, Green's Function Monte Carlo (GFMC) \cite{Pudliner1995,Carlson2015}, or Configuration-Interaction (CI) methods \cite{Sherrill1999}.
However, as a result of the many-body problem in physics (which observes the simple fact that the complexity of describing nuclear systems grows combinatorically with atomic mass number), it becomes quickly apparent that these and/or similar methods cannot be reasonably applied to even modestly heavy ($A\gtrsim 100$) nuclei, much less so for the heaviest/superheavy nuclei with $A\sim 300$; the computational resources that would be needed far exceed that which could be envisioned any time in the near or distant future.

Alternatively, energy density functional (EDF) theoretic methods in nuclear modeling intend to capture, in an approximate way, some of the rich microscopic physics present in atomic nuclei while avoiding the combinatorical explosion of complexity otherwise introduced in heavier nuclear systems \cite{Dobaczewski1984,Beiner1975,Bender2003,Stone2007,Goriely2009b,Schunck2019}.
These approaches focus on describing the nuclear potential energy in terms of local, one-body particle number densities and related quantities (kinetic energy density, spin densities, and spin-orbit densities, among a number of others).
Application of the variational principle immediately offers a path to identifying the ground-state configuration by minimizing the energy density functional with respect to all possible configurations, from which nuclear mass (among other nuclear structure observables) may be inferred.
Nuclear pairing effects prove something of a challenge for these methods, so model predictions in the nuclear EDF approach are generally restricted to even-even nuclei, although methods have been developed to also treat nuclear systems with unpaired protons and/or neutrons, e.g., via a blocking procedure \cite{Schunck2010}.
With phenomenological corrections applied on top of `bare' model predictions, these models can generally match known data on the order of a few hundred keV \cite{Goriely2009c,Kortelainen2010,Kortelainen2012,Erler2012,Goriely2013c,Agbemava2014,Kortelainen2014,Goriely2015c,Goriely2016}.
This situation is largely comparable to that of macroscopic-microscopic approaches; note, however, that in either approach, experimental measurements have characteristic uncertainties around an order of magnitude (or more) lower than any theory currently provides.

\begin{figure*}
 \centering
 \includegraphics[width=\textwidth]{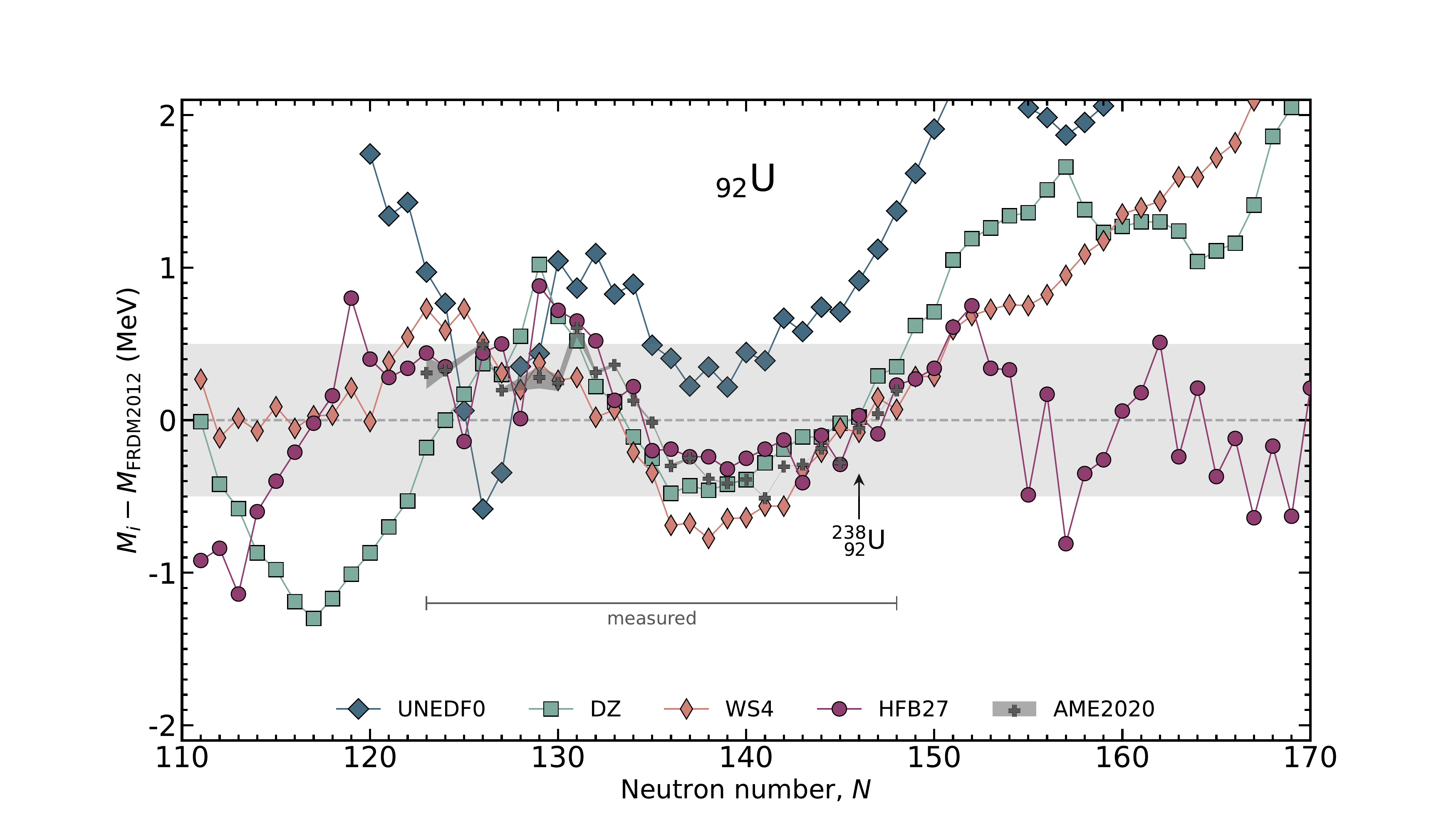}
 \caption{Mass model predictions along the U ($Z=92$) isotopic chain relative to FRDM2012. The plus markers and dark grey band show mass measurements and uncertainties, respectively, from each $Z=92$ entry in AME2020. (For clarity, the extent of current mass measurements along this isotopic chain is also indicated.) The light grey horizontal band indicates an overall 500 keV uncertainty, which is readily achieved by theoretical models within the region where mass measurements exists. Outside measurements, model predictions diverge rapidly.}
 \label{fig:mass_extrapolate_Z92}
\end{figure*}

In any of these approaches, theoretical mass models use experimentally known nuclei to constrain the underlying parameters within the model.
In some cases, other information can also be used in this parameter optimization step, including properties of extended nuclear matter, fission barrier heights, nuclear deformation, nuclear saturation density, the nucleon-nucleon interaction, and neutron skin depths.

Of chief concern for studying the production of the heaviest nuclei is how well models perform when extrapolating far from known, measured nuclei \cite{Pearson2006,Sobiczewski2014a,Sobiczewski2014b,Goriely2014}; open questions still persist around the evolution of deformation and collective effects \cite{Casten1987,Casten1999,Barbero2012}, as well as the strength of shell closures \cite{Strutinsky1967,Tondeur1978,Retamosa1997,Hirsch2010}.
Figure \ref{fig:mass_extrapolate_Z92} shows the behavior of several mass models for the uranium isotopic chain along with measurements from the AME2020 compilation where available.
All sets of masses are compared to a baseline mass model, FRDM2012, which allows for the opportunity to gauge extrapolations.
The difference between the mass model predictions and AME are within the standard uncertainty (500~keV) for theory mass calculations where there is measured data.
However, outside of the range of measured masses, the divergence between mass models quickly exceeds 500~keV, especially on the neutron-rich side.
This divergence of model predictions with increasing neutron excess is a general trend with current modeling efforts.

Accurate mass measurements of neutron-rich nuclei (those on the order of $\sim$25 keV uncertainty or less) are the primary way to resolve discrepancies among mass predictions \cite{Dilling2018}.
Measurements at modern facilities have provided a wealth of knowledge in recent years \cite{Knobel2016,Izzo2018,Ascher2019,Reiter2020,Smith2020,Li2022,Orford2022}.
Insights include the behavior of highly deformed nuclei \cite{Orford2018,Vilen2020,Vilen2020errat}, the behavior of nuclear near shell closures \cite{Baruah2008,Meisel2015,Lascar2017,Tang2020}, the resolution of isomeric states \cite{Yordanov2016,Ayet2019,Hornung2020} and the first direct mass measurement of a superheavy nuclide \cite{Schury2021}.

As we will see in the following subsections, the masses of nuclei play a critical role in nuclear reaction and decay rates, and remains one of the largest sources of uncertainty for the properties of neutron-rich nuclei.
Section \ref{sec:nucleosynthesis} will discuss in detail how different theory predictions for nuclear masses affect the production of the superheavy elements, helping or hindering the potential to observe them.

\subsection{Neutron capture}
\label{sec:neutron_capture}

Neutron capture is a transmutation process in which a nucleus absorbs a neutron, increasing the mass number by one unit.
This process is typically followed by the emission of $\gamma$-rays in the excited compound nucleus.
For the incident neutron energies ($E_{\rm n} \sim$ 1--100 keV) relevant to astrophysical processes that creates the heaviest elements, neutron capture can be described by Hauser-Feshbach theory \cite{Koning2005,Herman2007,Kawano2016}.
This statistical approach is based on the formation of a compound nucleus and is valid for nuclei that have sufficiently high level densities \cite{Kawano2006,Koning2008}.
For nuclei with low level density (e.g., near closed shells) or for nuclei near the neutron dripline, the direct capture mechanism can also contribute in a significant way \cite{Mathews1983,Rauscher1998,Goriely1997,Otsuki2010}.
Pre-compound reaction mechanisms such as semi-direct capture are less likely to play a role at the relatively low neutron energies found in astrophysical environments \cite{Xu2014}.

Statistical model calculations require inputs from nuclear structure models including the optical model potential \cite{Koning2003,Kunieda2007}, nuclear level density \cite{Demetriou2001,Goriely2008,Minato2011}, and the $\gamma$-strength function \cite{Bartholomew1973,Kopecky1990,Uhl1994,Liddick2016}.
The optical model---together with the associated optical model potential---is used to describe the overall cross section for the absorption of a neutron into a compound nucleus.
The nuclear level density and $\gamma$ strength functions are then used to infer the relative probability of the compound nucleus to de-excite.
When the excitation energy is well above the neutron separation energy, as in the case with $\beta$-decay of neutron-rich nuclei, neutron emission will dominate $\gamma$ emission in the statistical approach.
In contrast, for neutron capture in astrophysical environments, the total excitation energy in the compound nucleus is generally below the particle emission threshold, thereby limiting the reaction channels primarily to elastic, inelastic or radiative ($\gamma$) emission.
In radiative neutron capture, denoted (n,$\gamma$), the nucleus de-excites by the emission of one or more photons populating either the ground state or an isomeric state of the compound nucleus \cite{Mumpower2017}.
For some light nuclei the (n,$\alpha$) channel and its inverse may become energetically favorable \cite{Bliss2017, Psaltis2022} while neutron-induced reactions on nuclei near the neutron dripline \cite{Hansen2001} may also exhibit the (n,2n) channel due to the low separation energies \cite{Goriely2011}.
In both of these cases the channels will compete with (n,$\gamma$).
Besides the competition between neutron and $\gamma$ emission of the compound nucleus, for very heavy nuclei, the compound nucleus can be found in a state in which nuclear fission may also compete; this is discussed in much more detail in Section~\ref{sec:fission}.

Closely related to neutron capture is its inverse process, photodissociation, in which a photon has enough energy to liberate a neutron from the nucleus.
The energy required for such photons to cause photodissociation must naturally exceed that of a nucleus's one-neutron separation energy, without which the combined photon-nucleus system would have insufficient energy to strip a free neutron from the nucleus.
Unlike neutron-capture cross sections, which must generally be modeled directly, the rate for the inverse process may quickly be arrived at by applying arguments based on statistical mechanics---in particular, the principle of detailed balance.
In this way, the photodissociation rate can be expressed via the equation:
\begin{equation}
  \lambda_{(\gamma,n)} = \frac{2}{n_n} \frac{G'(T)}{G(T)} \left(\frac{\mu k_B T}{2\pi \hbar^2}\right)^{3/2}\lambda'_{(n,\gamma)} \mathrm{e}^{-S_n/k_BT},
\end{equation}
where $n_n$ is the neutron density; $G(T)$ is the nuclear partition function; $\mu$ is the reduced mass of the target and projectile nuclei; $T$ is the temperature; $S_n$ is the neutron separation energy; $\lambda_{(\gamma,n)}$ and $\lambda_{(n,\gamma)}$ are the photodissociation rate and neutron capture rate, respectively; and primed quantities correspond to those of the nucleus with one fewer neutron relative to those of the unprimed quantities.
For any fixed temperature, the separation energy controls the photodissociation rate insofar as it appears inside the exponential factor.

Overall, for a fixed neutron density, the specific point along an isotopic chain at which equilibrium is reached between the two processes, i.e., where the flow of nuclear abundance `stalls,' requires a detailed understanding of \textit{both} the astrophysical conditions (which sets the temperature of the nucleosynthetic environment) as well as nuclear structure for  neutron-rich nuclei across  the chart of nuclides (which sets the one-neutron separation energies).
The specific implications of this relationship for heavy-element nucleosynthesis are discussed further in Sec.\ \ref{sec:nucleosynthesis}.

\subsection{Nuclear decay}

A basic property of a nuclear species is how long it may persist, on average, before succumbing to radioactive decay.
The radioactive decay of a species involves changing the identity of one or more nucleons (as in the case of nuclear $\beta$ decay, Sec.\ \ref{sec:beta_decay}) or breaking into multiple, smaller nuclei ($\alpha$ decay and fission, Sec.\ \ref{sec:alpha_decay} and \ref{sec:fission}, respectively).
The elapsed time for a collection of identically prepared species to reduce to half of its initial population via radioactive decay is known as a half-life, $T_{1/2}$.
It is important to recognize that this definition is statistical in nature, and thus describes the average behavior of an ensemble of identically prepared states.

\begin{figure*}
 \centering
 \includegraphics[width=0.9\textwidth]{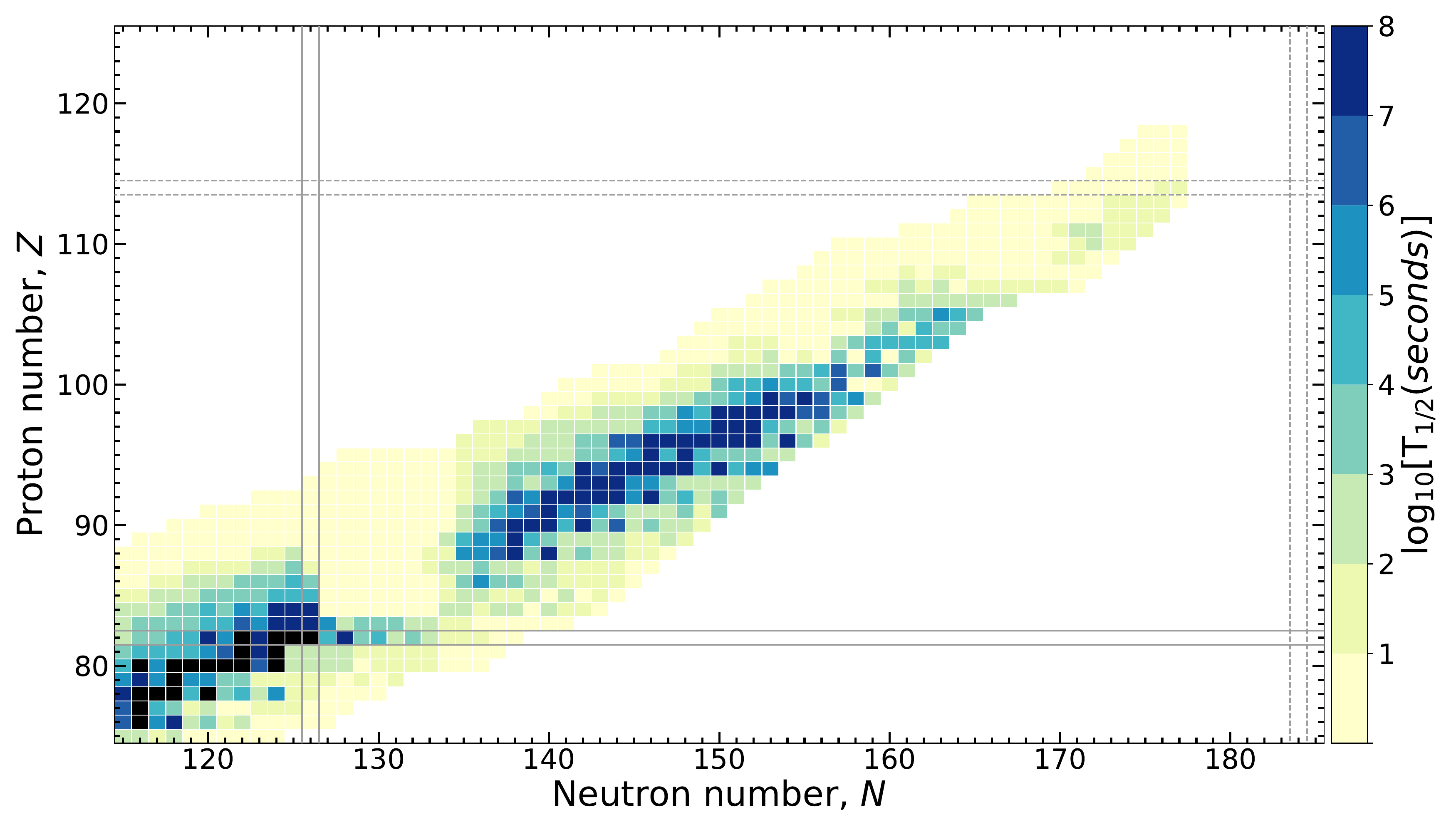}
 \caption{Half-lives of the heaviest measured isotopes. Note the extremely short half-lives encountered by species after $^{208}$Pb ($Z=82$) and before $^{223}$Ra ($Z=88$). Closed spherical shells that are confirmed (solid lines) and proposed (dashed lines) are shown for reference.}
 \label{fig:t12heavy}
\end{figure*}

Half-lives of the ground state of nuclei may range from fractions of a second to longer than the age of the universe.
Figure \ref{fig:t12heavy} shows the total half-lives for the heaviest measured isotopes as reported in the NuBase (2020) evaluation \cite{Kondev2020}.
Stable isotopes are indicated by black squares while radioactive nuclei are colored according to (the $\log_{10}$ of) their half-life.
Many elements in the actinide region, as well as all but two elements with a lower charge than bismuth ($Z<83$), have at least one long-lived isotope.
The exceptions of lighter elements with no stable isotopes are technetium ($Z=43$) and promethium ($Z=61$).
A gap of very short-lived nuclei exists between these two regions where $\alpha$-decay is the dominant decay mechanism.
Outside the region of measured masses, nuclei become evermore short-lived and, for lighter nuclei below $Z \sim 90$, are predicted to primarily undergo $\beta$-decay.

\begin{figure*}
 \centering
 \includegraphics[width=0.9\textwidth]{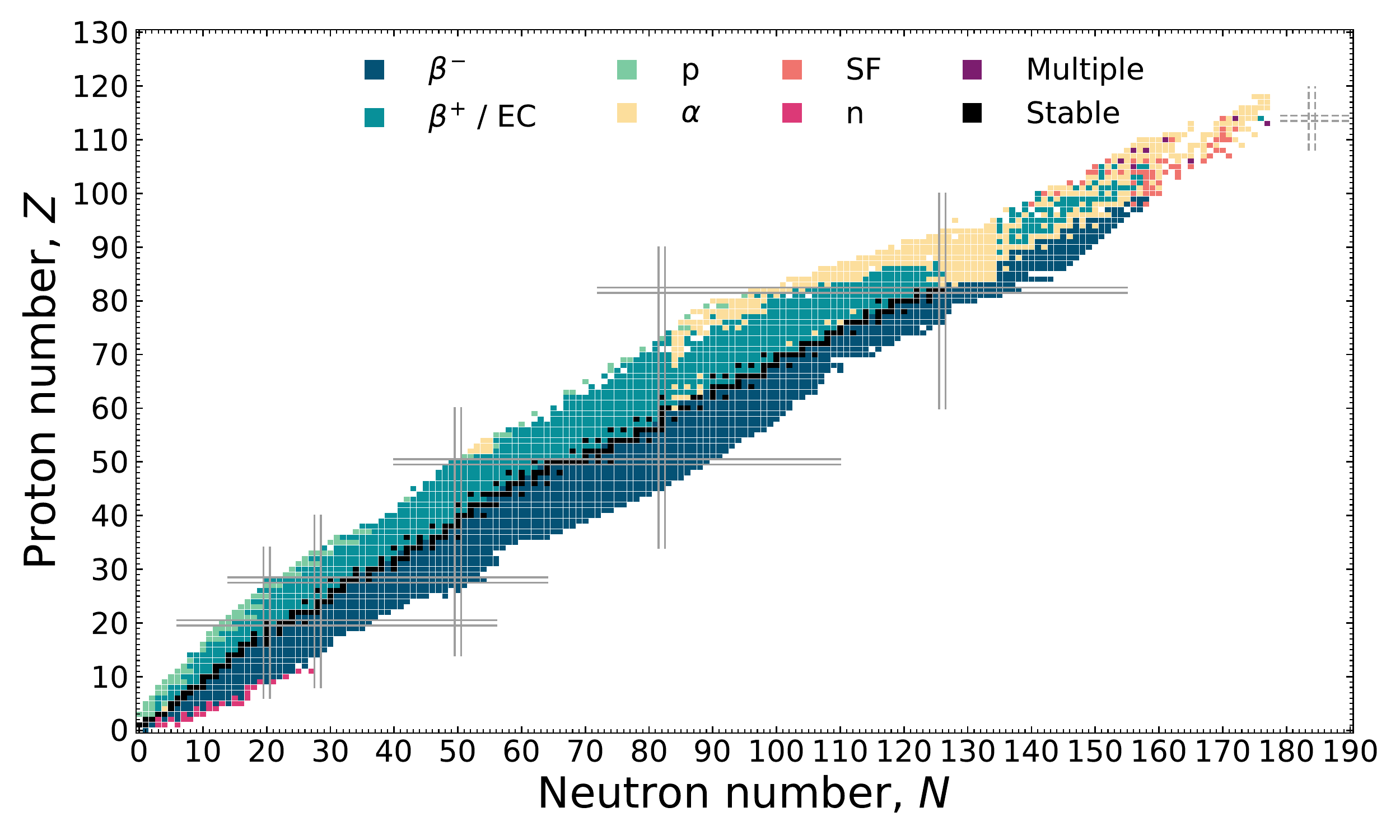}
 \caption{Primary measured decay mode across the chart of nuclides. Evaluated data taken from NuBase (2020) \cite{Kondev2020}. Closed shells that are confirmed (solid lines) and proposed (dashed lines) are shown for reference above $N/Z = 20$.}
 \label{fig:measured_decay_modes}
\end{figure*}

When two or more nuclear decay processes proceed in tandem, the total half-life is the sum of the inverse partial half-lives, which are defined as if each of the decay processes happened independently.
The total half-life is particularly relevant for the heaviest nuclear species where $\beta$ decay, $\alpha$ decay, and (spontaneous) fission may all compete.
Figure \ref{fig:measured_decay_modes} plots the primary measured decay mode across the chart of nuclides.
The majority of nuclei far from stability (including those involved in the creation of the heaviest elements) will undergo $\beta$ decay.
However, in the heaviest region, there is competition between these channels, and any individual nucleus may have significant chance to decay by $\beta$, $\alpha$, or fission.

Although the specific half-life of a nucleus depends on the combined probabilities of all its decay channels (which typically involve $Q$-value differences), the location of the \emph{most} $\beta$-stable nucleus along an isobaric chain can nonetheless be approximated from the binding energy.
Returning to the semi-empirical mass formula, one can estimate the minimum binding energy per nucleon (and therefore which combination of $Z+N=A$ is the most stable):
   \begin{equation}
   N / Z \sim 1 + \frac{a_C}{a_A}(Z+N)^{2/3},
   \end{equation}
where $a_{C}\sim0.7$ MeV and $a_{A}\sim23$ MeV are the Coulomb and asymmetry coefficients.
The small, positive factor implies that heavier stable nuclei have slightly more neutrons than protons.
This fact can be observed in Figs.\ \ref{fig:t12heavy} and \ref{fig:measured_decay_modes} as the known stable nuclei bend progressively away from $N=Z$.

Below, we discuss the principle modes of spontaneous decay for heavy, neutron-rich nuclei that involve particle emission. Spontaneous fission and $\beta$-delayed fission, which are also important in determining the overall (in)stability of the heaviest nuclei, are reserved specifically for our broader discussion of fission in Sec.\ \ref{sec:fission}.

\subsubsection{Beta ($\beta^{-}$) decay}
\label{sec:beta_decay}

Nuclear $\beta^{-}$ decay is the primary decay mode for the majority of nuclei involved in the neutron-capture processes responsible for creating the heaviest elements, the \emph{r} process.
This nuclear transmutation process is governed by the weak interaction that converts a neutron into a proton in the nucleus.
The nucleus simultaneously emits an electron and an anti-neutrino to conserve charge and lepton number.
In a somewhat humorous letter by Pauli \cite{Brown1978}, the neutrino itself was postulated in order to explain the continuous energy distribution of the electron that was observed decades earlier \cite{Sime1996}.
Later, Fermi provided a theoretical framework for performing calculations \cite{Fermi1934} (see also Wilson's English translation \cite{Wilson1968}).
For completeness---though not especially relevant during the astrophysical production of heavy nuclei via rapid neutron capture---$\beta^+$ decay occurs on the neutron-deficient side of nuclear stability when a proton in the nucleus transmutes into a neutron via the emission of a neutrino and a positron.
The decay of nuclei via $\beta^+$ (or $e^-$-capture) may contribute non-negligibly in the slow neutron-capture process, especially at branch points.
Unless otherwise specified, any mention of `$\beta$' decay here refers to $\beta^{-}$ decay.

The $\beta$-decay process may leave the nucleus in an energetically excited state.
In this excited state, the nucleus can either emit energy in the form of photons to decay to its ground state, or, if energetically allowed, emit ($\beta$-delayed) particles, possibly changing its identify yet again (a possibility we will discuss later in this section and in Sec.\ \ref{sec:fission}).
The total energy available in $\beta$ decay is measured by $Q_{\beta}$, the energy (or mass) difference between the initial and final ground states: $Q_{\beta} = m_{i} - m_{f} - m_{e}$.
Whether the daughter nucleus decays to a ground state or into a different nucleus depends on how much of the total available energy ($Q_\beta$) is allocated to the daughter nucleus.
Theoretical models of nuclear $\beta$ decay require both the relative energy between the parent and daughter states as well as a prediction of the transition strength (or intensity) between these participating levels \cite{Menendez2016}.
This requirement amounts to applying Fermi's Golden Rule to specific nuclear operators \cite{Engel2015,Engel2017,Nanni2018}.
The pairing of this information may be summarized in a $\beta$-decay strength function of a nucleus \cite{Hansen1973}, from which the overall $\beta$ decay rate may be immediately calculated \cite{Moller1990}.
When, following $\beta$ decay into an excited state, the population of excited states in the daughter nucleus exceeds particle threshold, delayed particle emission may ensue \cite{Pappas1972,Gjotterud1978}.
For neutron-rich nuclei, delayed particle emission may occur when the neutron separation energy in the daughter nucleus is less than the parent's $Q_{\beta}$, i.e., $S_n (Z+1,N-1) < Q_{\beta}(Z,N)$, leading to so-called $\beta$-delayed neutron emission \cite{Borzov2014,Dimitriou2021}.

Overall, the excited nucleus will favor particle emission over electromagnetic decay whenever it is energetically allowed \cite{Spyrou2016}.
However, a more refined portrait may be obtained by employing principles of statistical decay, in very much the same way as de-excitation is modeled following the formation of a compound nucleus \cite{bohr1935} following neutron capture (Sec.\ \ref{sec:neutron_capture}).
In this way, the branching ratios and energy spectra of emitted particles may be calculated \cite{Kawano2008,Mumpower2016b}.
Because the overall energy of the compound nucleus is determined by $Q_\beta$, which can be quite large (ten MeV or more) for the very neutron-rich nuclei populated during an \emph{r} process, it is energetically favorable for multiple neutrons to be emitted before a $\beta$-unstable nucleus finds its way to a stable nucleus.
For nuclei with large $Q_\beta$, i.e., as one approaches the neutron dripline, modern models predict approximately five delayed neutrons emitted per $\beta$ decay \cite{Minato2021}.
In the most extreme cases, as many as eight or more neutrons may be emitted following $\beta$ decay in \emph{r} process nuclei.

\begin{figure*}
 \centering
 \includegraphics[width=0.85\textwidth]{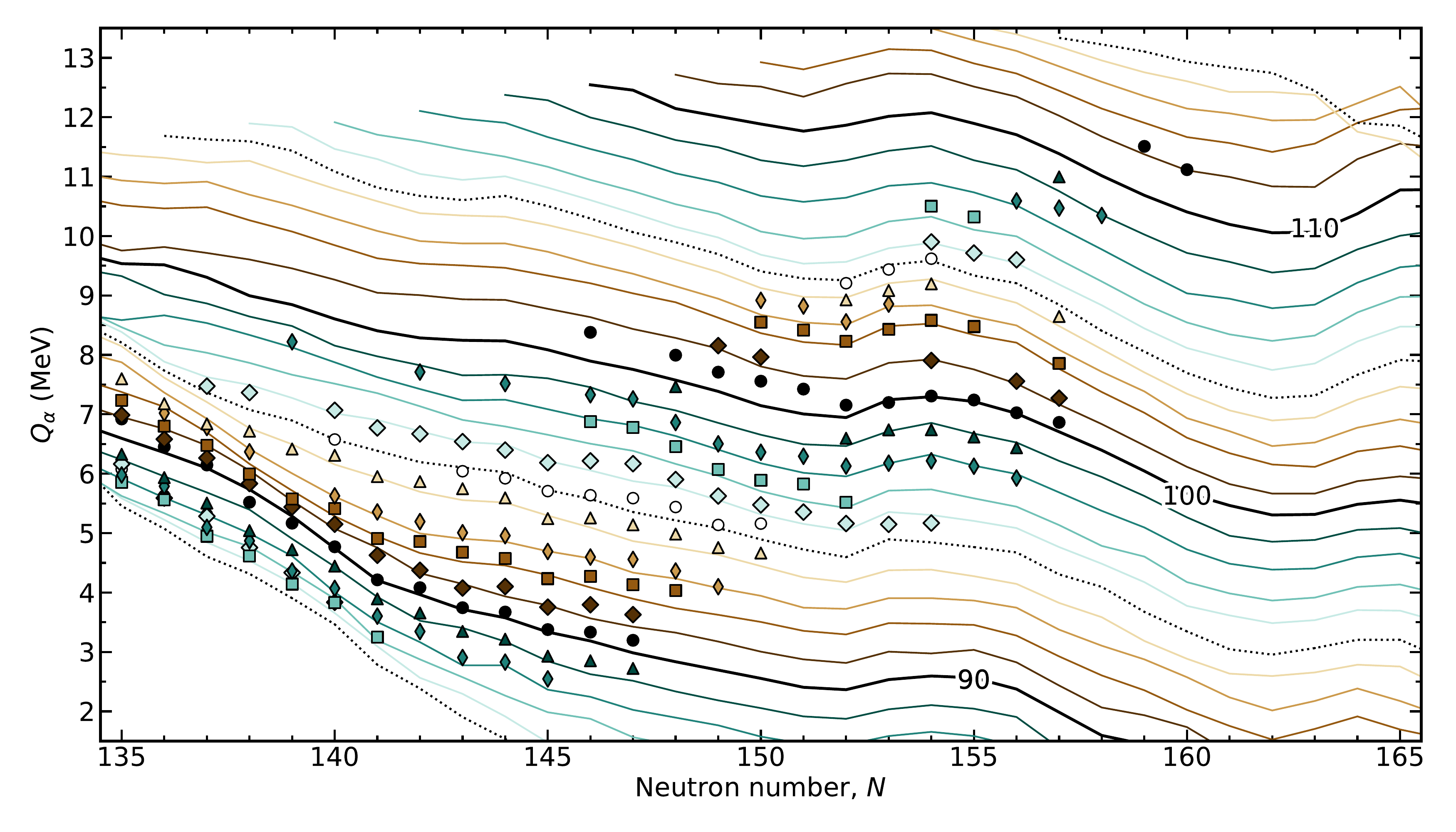}
 \caption{Isotopic $\alpha$-decay $Q$-values from the 2020 release of the Atomic Mass Evaluation (AME2020) (points) \cite{Wang2021} compared with those from the FRDM2012 model (solid lines) \cite{Moller2016} along constant proton number. Thick black lines indicate proton-number intervals of ten.}
 \label{fig:qalpha}
\end{figure*}

For nuclei in a localized region of the chart of nuclides, the nuclear shell model may be used to calculate $\beta$-decay rates \cite{Langanke2000,MartinezPinedo1999,MartinezPinedo2000,Zhi2013}.
However, currently, to compute all relevant nuclei, theories with a global reach must be employed \cite{Minato2013,Minato2022}.
Global calculations of $\beta$-decay properties have been calculated within the macroscopic-microscopic method based on the Quasi-particle Random Phase Approximation (QRPA) \cite{Moller1997} with phenomenological treatment of first-forbidden transitions \cite{Moller2003}.
This approach generally yields $\beta$-decay properties dominated by Gamow-Teller transitions \cite{Moller2019}.
In contrast, microscopic approaches suggest that first-forbidden transitions play a strong role, especially above the $N=126$ shell closure \cite{Marketin2016}.
More recently, microscopic approaches have utilized the finite-amplitude method to speed up the calculation of $\beta$-decay properties \cite{Mustonen2016,Ney2020}.
It is important to keep in mind that the current large discrepancies in $\beta$-decay properties will have substantial impact on the creation of the elements in astrophysical environments.
We return to this point in Section~\ref{sec:nucleosynthesis}.

\subsubsection{Alpha ($\alpha$) Decay}
\label{sec:alpha_decay}

Alpha decay is a nuclear transmutation in which an $\alpha$ particle (a helium nucleus consisting of two protons and two neutrons) is emitted from the nucleus.
Alpha decay is technically possible when the $Q$-value ($Q_{\alpha} = m_{i} - m_{f} - m_{\alpha}$) is positive, and it is one of the primary decay modes of the heaviest nuclides.
However, because of a relatively strong Coulomb barrier established by the parent nucleus, there are many nuclei which exhibit a positive $Q_\alpha$ but which undergo $\alpha$ decay with effectively zero probability, and those nuclei with significant $\alpha$ decay rates are those with comparatively larger $Q_\alpha$ values.

\begin{figure*}
 \centering
 \includegraphics[width=0.85\textwidth]{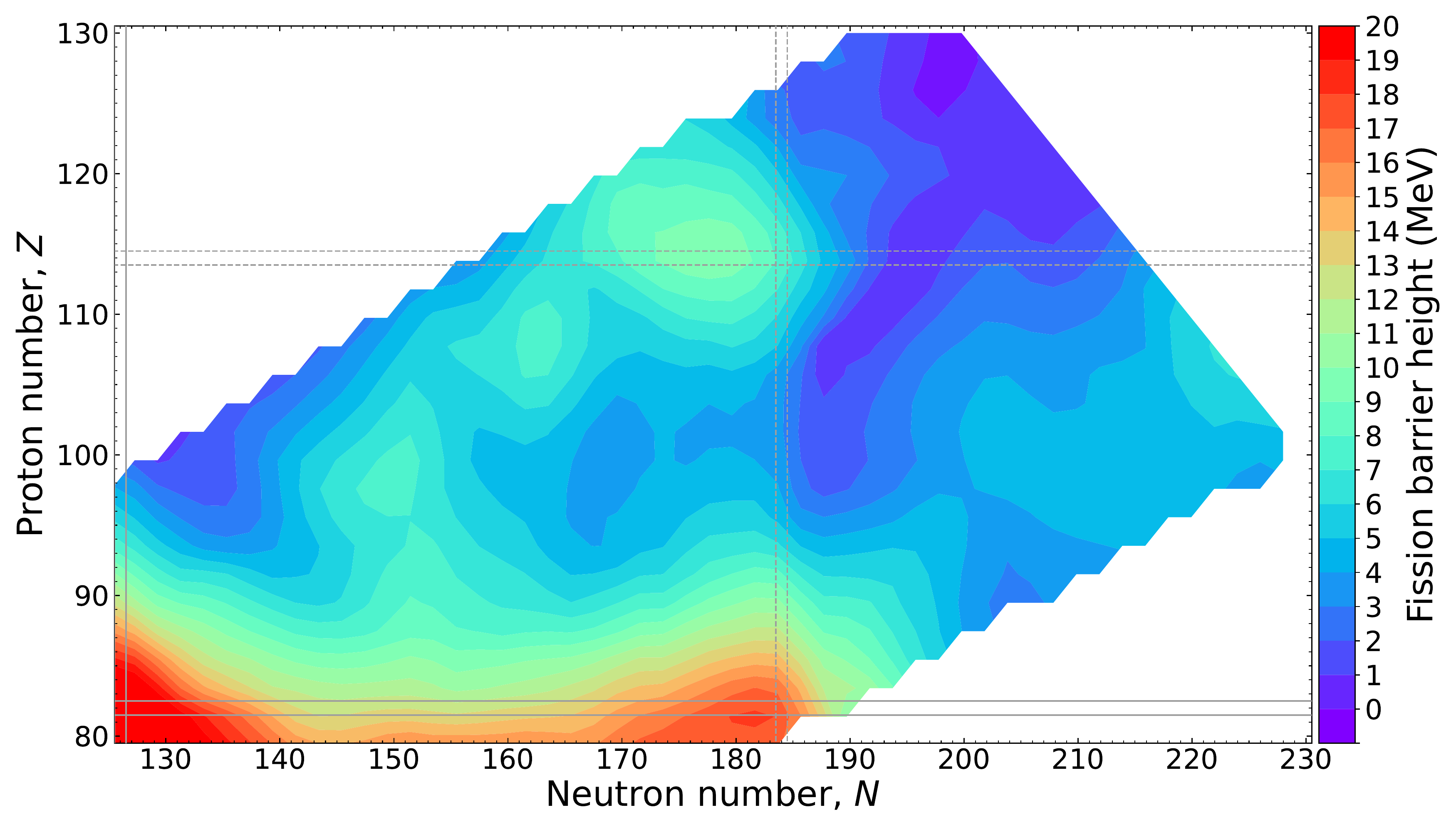}
\caption{Fission barrier heights across the heaviest region of the chart of nuclides, as calculated using the FRLDM model \cite{Moller2015}. Note the region of particularly low barrier heights around $Z=100$ and $N=190$, which boosts nuclei's tendency towards fission and substantially inhibiting their capacity to produce the superheavy elements.}
\label{fig:barrier}
\end{figure*}

Figure \ref{fig:qalpha} shows the behavior of $Q_{\alpha}$ along isotopic chains for heavy nuclei.
Measurements reported in the AME2020 are shown in points while lines indicate $Q_{\alpha}$ predictions from FRDM2012.
Black solid (dotted) lines specify proton numbers divisible by ten (five).
This figure reflects the general qualitative agreement between theory and experimental values when used to determine relevant $Q$-values for nuclear reaction and decay calculations.
In reference to the \emph{r} process specifically, however, it should be noted that those nuclei with significant $Q_\alpha$ lie much closer to those with known masses, so the difficult and persistent complication of uncertain nuclear masses in relation to, e.g., theory calculations of neutron capture or $\beta$-decay of very neutron-rich nuclei is not nearly so severe for analogous predictions of $\alpha$-decay properties.

The partial half-life of $\alpha$-decay may be calculated via an assumption of quantum tunneling through a barrier, which results in the well-known Viola-Seaborg relation \cite{Viola1966}.
This kind of phenomenology was crucial in anticipating the possible $N=184$ shell closure.
Following this work, subsequent predictions of $\alpha$-decay half-lives have been made using various methods \cite{Poenaru1983,Samanta2007,Kiren2013,Sahu2016,Santhosh2017,Santhosh2018}.
While measurements have recently been made of lighter $\alpha$-decaying nuclei \cite{Braun2017}, it is the $\alpha$-decay chains of the heaviest nuclei that impact the actinide and superheavy elements.

\subsection{Nuclear fission}

The heaviest nuclear species may undergo fission, which is the general term for processes that splits the nucleus into two or more lighter `fission fragments.' 
The splitting of a nucleus into two main fragments along with, possibly, some small handful of neutrons is the most common modality, as first reported by Meitner, Strassman, and Hahn \cite{Meitner1938}.

Fission rates are the most important fission property in controlling \textit{whether} and \textit{how many} heavy and superheavy nuclei might be produced following an \emph{r} process; if the rates for certain nuclei are too fast, for example, nuclei will have a small `survival probability' during their decay towards less neutron rich nuclei that occurs at the end of any \emph{r} process.
At the same time, electromagnetic and particle radiation spectra, as well as fission fragment yields, can be important observational tools for examining the possibility of heavy and superheavy element production in \emph{r}-process environments and their associated remnants.

Overall, a complete theoretical framework for calculating the numerous fission properties of interest to the \emph{r} process has proven to be a formidable task, and a number of approaches have been developed to make progress towards this persistent challenge in nuclear theory.
Broadly speaking, the theoretical approaches mirror those applied in nuclear structure (see discussion of Sec.~\ref{sec:masses}), with fully microscopic, macroscopic-microscopic, and phenomenological models all providing important developments in our understanding of, and predictions for, fission properties across the nuclear chart (see  Ref.~\cite{Schunck2022} for a recent and thorough review).
Microscopic approaches, in particular, are at a uniquely promising point in their history, as recent advances in computational hardware are beginning to provide the considerable resources needed to fully apply such methods to the problem of fission~\cite{Jach2017,Hessberger2017,Regnier2017,RodriguezGuzman2017,Sekizawa2017,Tao2017,Verriere2017,Giuliani2018,Navarro2018,Lemaitre2019,Zhao2019,Bulgac2020,RodriguezGuzman2020,Bulgac2021,Jin2021,Marevic2021, Verriere2021b,Bulgac2022} (also see Ref.~\cite{Schunck2019} and references therein). At the same time, macroscopic-microscopic approaches have continued to fill a gap insofar as their comparatively less demanding computational methods enable studies that predict fission properties across the entire nuclear chart \cite{Abe1996,Moller2004, Moller2015b,Randrup2011a,Randrup2011b,Sierk2017,Mumpower2020,Verriere2021}.
Still too, systematics fill important gaps in our ability to make reasonable predictions for certain types of fission observables, particularly with regards to the calculation of fission rates \cite{Hill1953,Bjornholm1980,Xu2005,Karpov2012,Panov2013} and approximate fragment yields \cite{Kodama1975,Wahl2002}.
Modern treatments now also include fragment energy partitioning, deexcitation, and corresponding neutron and gamma emission spectra \cite{Schmidt2016,Jaffke2018,Okumura2018,Lovell2021,Talou2021,Vogt2021}.

We explore the specific implications of fission for nucleosynthesis in Sec.\ \ref{sec:nucleosynthesis} while focusing the remaining portions of the present discussion on the modern understanding of fission in and of itself.


\subsubsection{Fission barriers}
\label{sec:fiss_bar}

In any prediction of fission rates---regardless of fission process or nucleus---the fission barrier is one of the strongest predictors of a nucleus's propensity to fission.
Because the fission properties of so few nuclei have been measured experimentally (and fewer still, when only nuclei of interest to the \emph{r} process are considered), \emph{r} process simulations rely on global models for predicting fission barriers which are, in turn, used to calculate actual fission rates for the relevant fission processes. Such global predictions are available in the BCPM \cite{Giuliani2013}, ETFSI \cite{Aboussir1992,Aboussir1995}, FRLDM \cite{Moller2015}, HFB-14 \cite{Arnould2006}, and Thomas-Fermi \cite{Myers1999} models.
In each of these models, the nuclear potential energy of each nucleus is computed across a domain of collective coordinates, and the resulting geometry of the potential energy surface is analyzed to identify the fission barrier in reference to the ground state and splitting (scission) configurations.
The difference between the ground state and fission barrier energies is taken to be the `fission barrier height,' which is the quantity ultimately used to predict the rates for the different fission processes \cite{Nix1972}.
When a nucleus is initially in an excited state (e.g., as in the compound nucleus formed in neutron-capture or the excited energy states nuclei $\beta$ decay into), the excitation energy may be subtracted from the barrier height to obtain an effective value for further use in such calculations.

In Fig.\ \ref{fig:barrier}, we show the global behavior of barrier heights across the fissioning region of the nuclear chart for the FRLDM model \cite{Moller2015}.
Particularly noteworthy is the transition to a broad, diffuse region of low barrier heights for very neutron-rich nuclei located around $Z=100$ and $N=190$.
In the 1960s, it was already expected that this general region would have relatively lower barrier heights that would enhance the propensity to fission as \emph{r} process attempt to traverse this region on a path towards the superheavy elements, ultimately inhibiting or altogether preventing the production of these elements in nature \cite{Moller2010}.
Modern calculations involving the aforementioned models generally agree with each other on the order of $5$ MeV (but diverge more significantly for the most neutron-rich nuclei), but in any case generally demonstrate this region of lower barrier heights, along with the corresponding implications for superheavy element production \cite{Thielemann1983,Giuliani2013,Mumpower2018,Vassh2019,Minato2021}.
Nevertheless, these calculations are subject to the same level of uncertainties characteristic of nuclear theory in general; future work that aims to improve on these methods is surely warranted.
We will explore in Sec.\ \ref{sec:production} how this region of low fission barrier heights affects superheavy-element formation in modern \emph{r}-process simulations.

\subsubsection{Fission Processes}
\label{sec:fission}

In the most general sense, fission occurs when a nucleus proceeds through a complex configuration space, temporarily passing through one or more regions associated with a relatively high potential energy, until it arrives at one of many possible highly deformed `scission configurations' with much lower potential energy than it began with.
Once a scission configuration is realized by the nucleus, the nucleus is energetically driven to split into two or more fragments.
The fission barrier measures the minimum excitation energy (relative to the ground state) that the nucleus must possess in its initial state to undergo this process in the classical sense, i.e., such that the total energy never drops below zero throughout its dynamical trajectory to fission.
Of course, the nucleus is clearly a quantum mechanical system, so it is often the case that the nucleus proceeds through the barrier by tunneling on occasion, and the fission barrier is therefore used as a proxy for predicting fission rates in phenomenological approaches.

The different fission processes observed in nature, then, are categorized in reference to the different ways that the initial state of a fissioning nucleus might be prepared.
Perhaps the simplest among these is the case of spontaneous fission, which can occur in ground or metastable states of especially heavy nuclei, in which fission proceeds \textit{in situ} by tunneling through the fission barrier \cite{Nilsson1968}.
In this case, the rate of spontaneous fission is a constant property of the nucleus (i.e., not determined by extrinsic factors such as temperature, density, or other nuclei) \cite{Randrup1973,Randrup1974,Randrup1976}.
Even though spontaneous fission may technically be possible for a large number of unstable heavy nuclei, it is frequently the case that it is negligible when compared with other available decay modes that proceed much more quickly.
For nuclei produced during the \emph{r} process, this is generally the case.
However, on longer timescales after the \emph{r} process has concluded, it is likely to contribute significantly once other decay and/or fission modes have become inaccessible (e.g., after the exhaustion of free neutrons effectively closes off the neutron-induced fission channel, discussed below).
While a small number of spontaneous fission rates have been experimentally measured \cite{Kondev2020}, the intersection of these instances with \emph{r} process nuclei is quite small.
Theoretical calculations may be based on barrier penetration arguments \cite{Sadhukhan2020a} or empirical phenomenology \cite{Xu2005}, some of which broadly capture the dependency of the rate on the fission barrier \cite{Zagrebaev2011,Karpov2012}.

In contrast with spontaneous fission, there are any number of conceivable ways that a nucleus can be prepared with a wide range of excitation energies.
When this occurs, a nucleus's total energy will be brought much closer to the fission barrier, which can dramatically increase the probability of fission.
Of particular interest to \emph{r} process applications are the neutron-induced and $\beta$-delayed fission processes, both of which may provide a nucleus with such excitation.

In neutron-induced fission, a compound nucleus forms in much the same way that of neutron capture.
The total energy of a nucleus immediately after it captures a neutron is equal to the sum of the neutron energy and the one-neutron separation energy ($S_n$) of the product nucleus.
In ordinary neutron capture, the compound nucleus will undergo a cascade of electromagnetic decays to the ground state.
However, for the heavier nuclei, where the fission barriers are sufficiently low, fission competes with the electromagnetic decays, such that some (possibly significant) fraction of neutron-capture events terminate with fission, rather than with the ground state associated with the compound nucleus.
The total neutron-induced fission and neutron-capture rates can be evaluated by considering both channels in statistical Hauser-Feshbach approaches \cite{Kawano2021}.

\begin{figure}[t]
 \centering
 \includegraphics[width=\columnwidth]{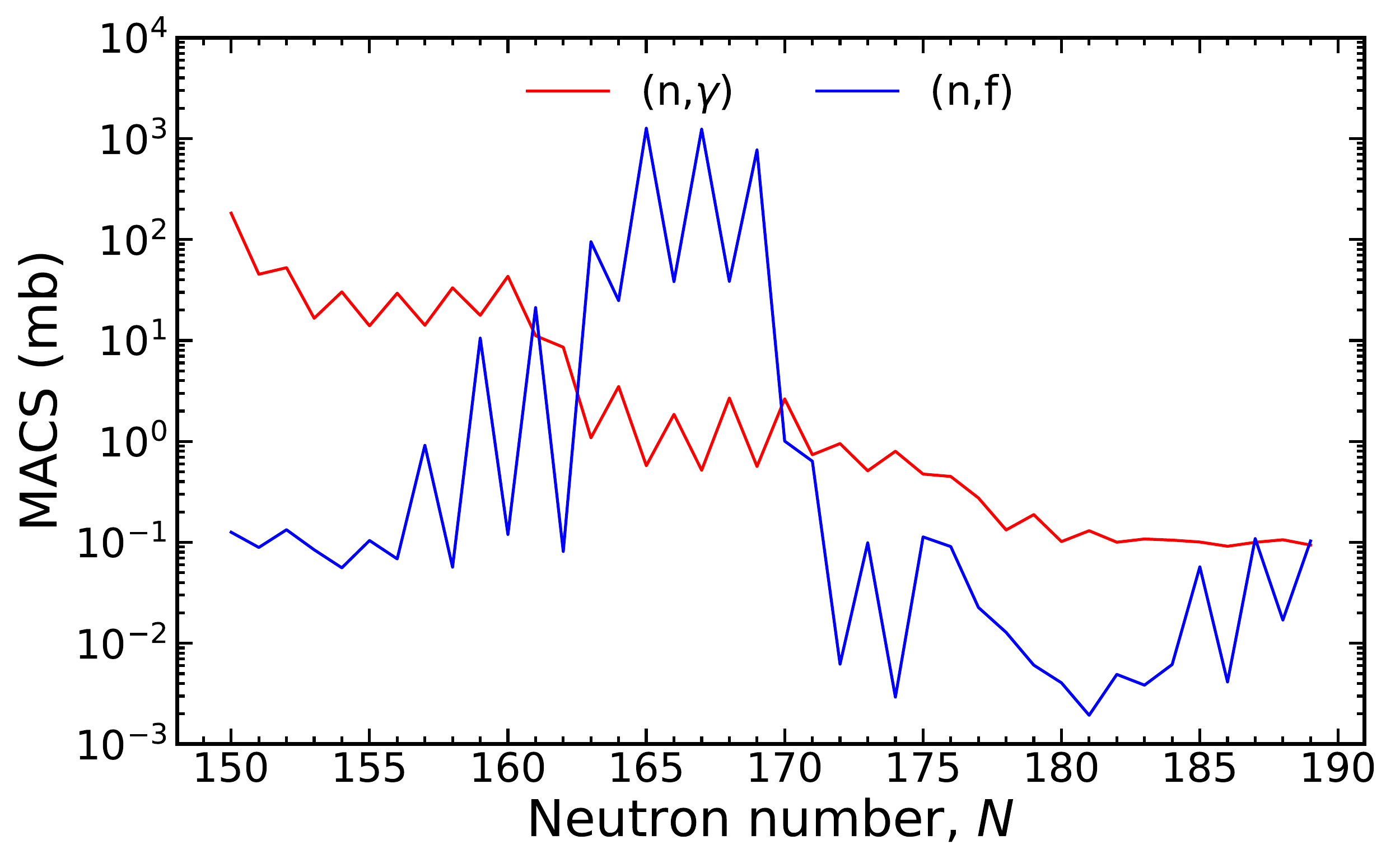}
\caption{Maxwellian averaged cross sections (MACS) for neutron capture (red) and neutron-induced fission (blue) across the Np ($Z=93$) isotopic chain. The cross sections are evaluated at $T_9=1$ GK, a typical temperature that an \emph{r} process might traverse through.}
\label{fig:macs}
\end{figure}

In Fig.\ \ref{fig:macs}, we plot both of the neutron capture and neutron-induced fission cross sections as a function of increasing neutron number for the Np ($Z=93$) isotopic chain and for thermalized environments at a temperature broadly representative of the \emph{r} process, $T=1$\,GK.
Recalling the fission barriers plotted in Fig.\ \ref{fig:barrier}, there is a diffuse region of considerably lower fission barriers that the $Z=93$ isotopic chain passes through for isotopes with neutron numbers roughly between $N=165$ and $N=175$.
This trend in fission barrier height is clearly reflected when comparing the relative cross sections for fission vs.\ neutron capture, where the sharp peak in the fission cross section precisely coincides with the nuclei with the lowest barriers, with neutron-induced fission clearly dominating over neutron capture by around three orders of magnitude in the most extreme instances.

Importantly, the total neutron-induced fission rate for each nucleus, $\lambda_{(n,f)}$, depends directly on free neutron number density, $n_n,$ such that $\lambda_{(n,f)} \propto n_n$.
Correspondingly, in the extremely neutron-rich environments characteristic of the \emph{r} process, neutron-induced fission can be quite significant, especially during the earlier phase of the \emph{r} process where the total number of neutrons exceed all other nuclei in the environment.
Indeed, network calculations suggest that neutron-induced fission can, for a brief but critical period of time during nucleosynthesis, be the fastest fission process \cite{Petermann2012}.

The third and final fission process typically considered in \emph{r}-process studies is $\beta$-delayed fission.
In this scenario, a nucleus undergoes an initial $\beta$ decay which leaves it in any number of excited nuclear states of the final nucleus, ranging from the ground state up to the $Q$-value for the decay, with the remaining energy being carried away by the electron and anti-electron-neutrino that are immediately produced by the decay.
As in neutron-induced fission, the excited nucleus formed immediately after $\beta$ decay has some probability to fission, the extent of which is determined largely by the relative magnitude of the excitation energy and the fission barrier.
The total $\beta$-delayed fission rate is then taken to be product of the total $\beta$-decay rate and the probability to subsequently fission, and the effective $\beta$ decay rate is the difference of the total rate with the $\beta$-delayed fission rate.

An interesting variation on $\beta$-delayed fission happens when the most neutron-rich nuclei of the \emph{r} process are considered.
As one moves across an isotopic chain towards increasingly neutron-rich nuclei, the neutron separation energy \textit{decreases} while the $Q$-value for $\beta$ decay \textit{increases}.
As discussed in Sec.\ \ref{sec:beta_decay}, this provides ideal conditions for the emission of one or more delayed neutrons.
By coupling a probability to fission at each stage of this decay sequence, and provided that the fission barriers are sufficiently low that fission becomes an appreciable possibility, this introduces the prospect for \textit{multi-chance} $\beta$-delayed fission, a process which comprises a spectrum of possibilities for the emission of some number of delayed neutrons followed by fission (as opposed to electromagnetic decay to a ground or metastable state).
The overall process of $\beta$ decay is then a complex convolution of electromagnetic decays, neutron emission, and fission.
The cumulative effects of the competition between these decay modes can be carefully resolved in a statistical Hauser-Feshbach framework \cite{Mumpower2018,Minato2021} in much the same way that the competition between different exit channels are treated to resolve the probability of neutron-induced fission relative to neutron capture.

While neutron-induced fission can be the fastest fission process during the first second(s) of an \emph{r} process, when free neutrons are still extremely abundant, $\beta$-delayed fission and multi-chance $\beta$-delayed fission rates, which are constant and necessarily \textit{do not} depend on neutron abundance, can become the primary fission process once the neutron abundance drops to sufficiently low levels, thereby eliminating any significant possibility for neutron-induced fission to occur.
Because this generally occurs when nuclei are decaying towards longer-lived and less neutron-rich nuclei, accurate predictions for $\beta$-delayed fission rates across the nuclear chart are especially important for understanding the formation of heavy and superheavy elements in the aftermath of the \emph{r} process \cite{Kodama1975,Thielemann1983,Petermann2012,Goriely2015b, Ghys2015,Mumpower2018,Holmbeck2019a}.

\section{Astrophysical alchemy}\label{sec:nucleosynthesis}

\subsection{The basic \emph{r} process}

\begin{figure*}
 \centering
 \includegraphics[width=0.85\textwidth]{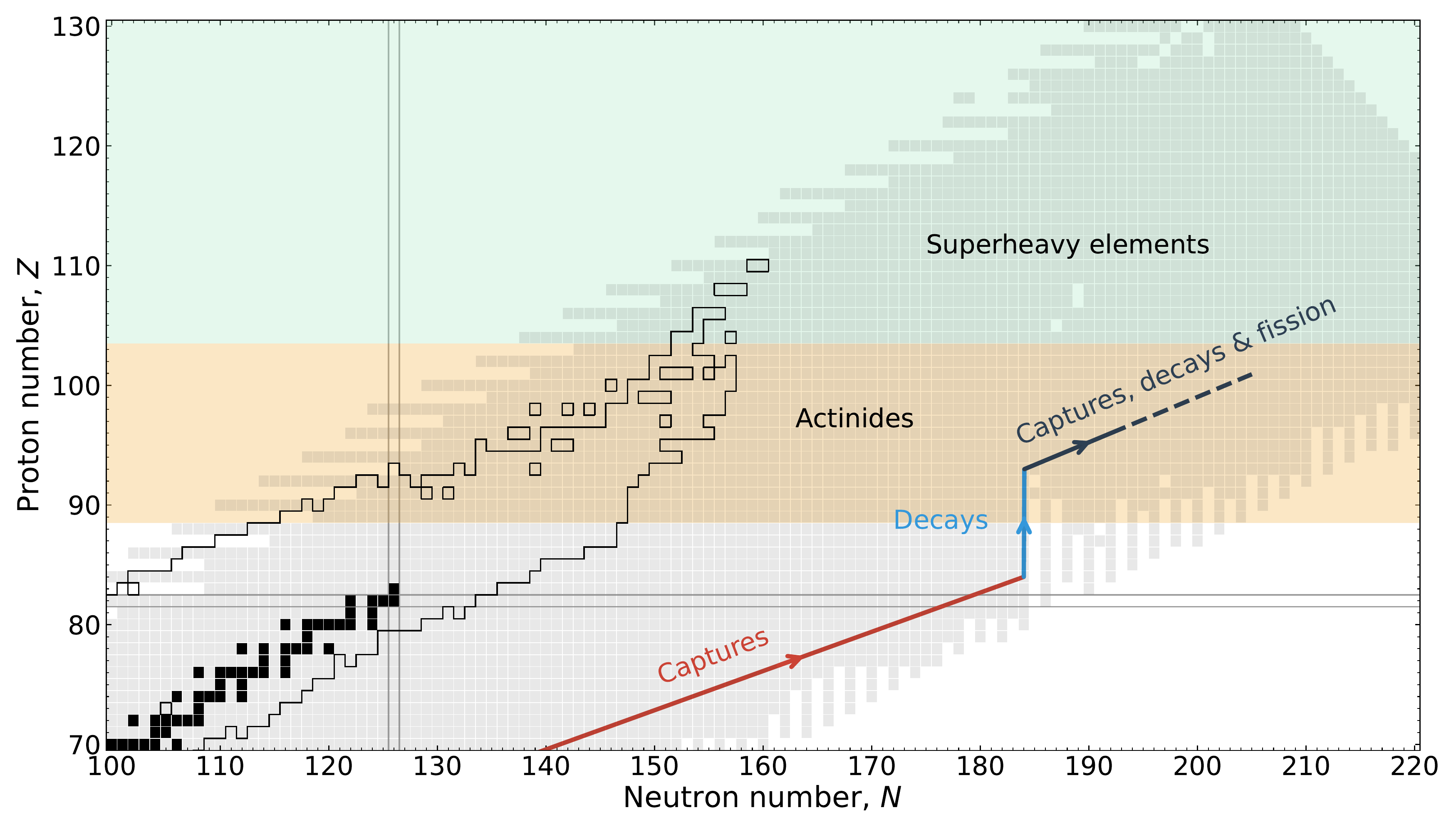}
\caption{Schematic description of the \emph{r}-process pathway and the dominant reaction channels active in the heavy-element region. Black boxes indicate stable nuclei, outlined boxes measured nuclei, and grey-filled a theoretical extent for which nuclei could possibly exist. Note that for these heavy-element regions, the \emph{r}-process pathway is very far from the extent of measured nuclear data (i.e., far from the black-outlined boxes denoting the extent of the AME2020), indicating that the nuclear data in these regions rely on theory and become very uncertain.}
\label{fig:rpath}
\end{figure*}

As we alluded to in Sec.\ \ref{sec:intro:nucleosynthesis}, the natural abundance of elements heavier than iron are expected to be mostly produced by neutron capture nucleosynthesis, with the heaviest among these (those heavier than lead) in particular being produced entirely by a subset of neutron-capture nucleosynthesis known as the \textit{rapid neutron capture process}, or the \emph{r} process \cite{Horowitz2019}.
Up until now, we have made many references to the \emph{r} process while describing the properties of the heaviest nuclei that are created---even if momentarily---by this nucleosynthetic mechanism.
Now, equipped with knowledge of the relevant nuclear physics in the heavy-element region, we may turn to a comprehensive description of the \emph{r} process itself.

Schematically, the \emph{r} process proceeds as follows: given a system with a small abundance of `seed' nuclei but subjected to an extraordinary flux of free neutrons, the neutrons will radiatively capture onto the seed nuclei, transmuting them along isotopic chains to nuclei with heavier mass numbers but the same atomic number.
In this way, the nuclei may be pictured as moving horizontally on the chart of nuclides (when plotted in $N$ versus $Z$).
As neutron number increases for a given isotopic chain, the $\beta$-decay rates tend to become much faster \cite{Shafer2016}.
At the same time, the neutron separation energies tend to \textit{decrease} as nuclei become more neutron-rich, such that the rate of neutron capture will become slower overall, with the inverse process, photodissociation of neutrons from nuclei, further reducing the overall capacity of nuclei to net-capture neutrons.
The combined effect of the increasing $\beta$-decay rate and decreasing neutron-capture rate is that the seed nuclei will continue to capture neutrons (forming heavier nuclei) until $\beta$ decay rates become fast enough to compete with neutron capture, which will then increment the nuclei to the next successive element.
This process repeats for the next isotopic chain, with neutron capture increasing mass number and $\beta$ decay transmuting nuclei into heavier elements.

Figure \ref{fig:rpath} shows a schematic for how the heaviest nuclei are populated during the \emph{r} process in terms of their primary reaction channels: captures, decays, and fission; recall discussion in Sec.\ \ref{sec:nucphys}.
The \emph{r}-process `path' shown by the lines with arrows approximately denotes a neutron-rich pathway near the neutron dripline traversed in creation of the heaviest elements.
Along the red line, there are more captures per decay, as indicated by its shallow slope.
The blue line indicates that the process of moving up a closed neutron shell---such as that predicted at $N=184$---is moderated by decays.
Once the predicted shell closure is surmounted, large uncertainties make the definition of the \emph{r}-process path (and the dominant channel) difficult to determine.

The \emph{r}-process path denoted in Fig.\ \ref{fig:rpath} is purely schematic.
It is more rigorously defined as the set of the most abundant nuclei along each isotopic chain at a given time in the \emph{r} process.
In the early stages of the \emph{r} process at high temperatures and densities, the path can be defined by an equilibrium between (n,$\gamma$) and ($\gamma$,n) along any given isotopic chain.
Once the bulk of these reactions are no longer in equilibrium, a point known as (n,$\gamma$) freezeout, the path may be defined by a subset of nuclei for which the $\beta$-decay rates have become competitive with the neutron-capture rates \cite{Cowan1991}.

\begin{figure*}
 \centering
 \includegraphics[width=140mm]{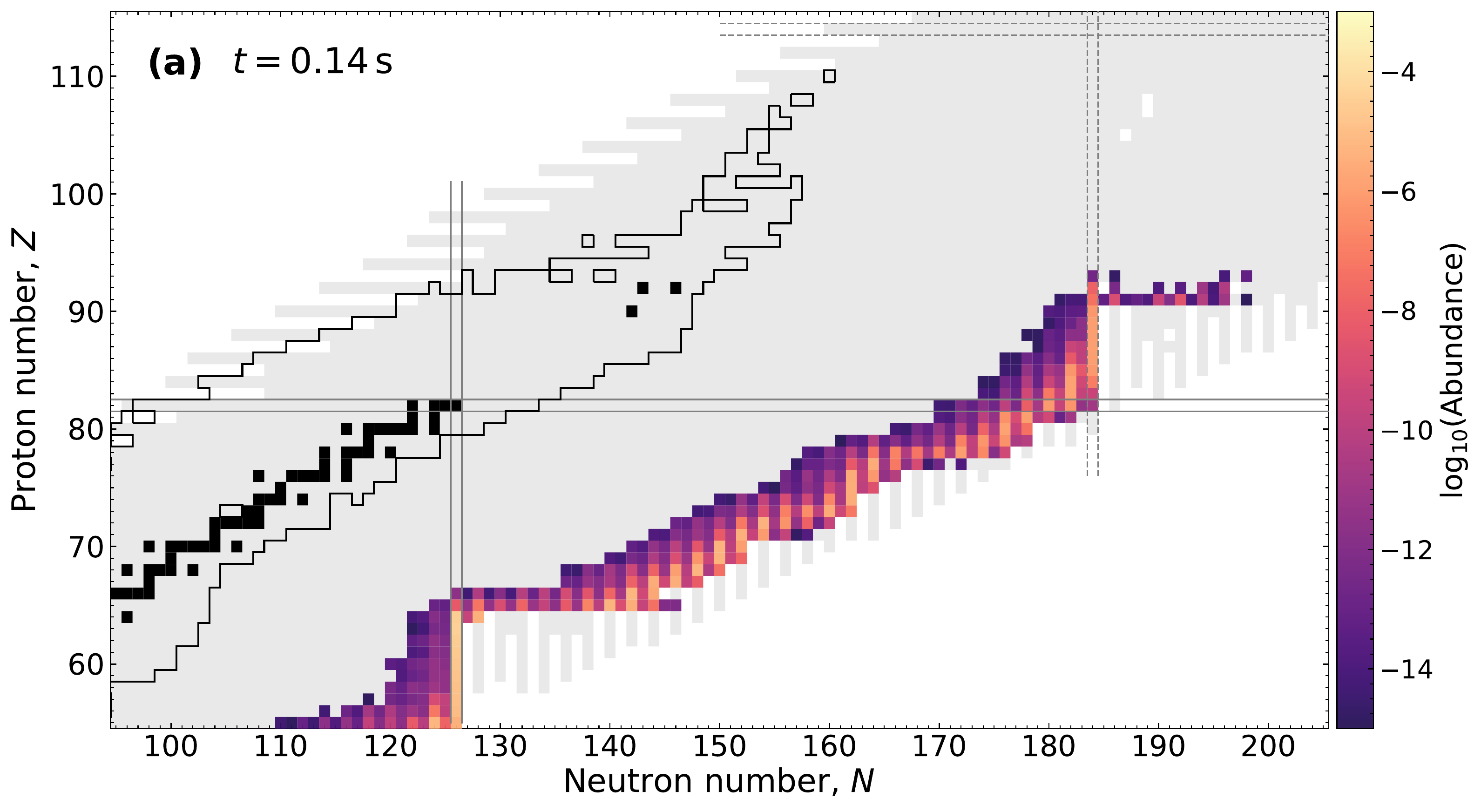}
 \includegraphics[width=140mm]{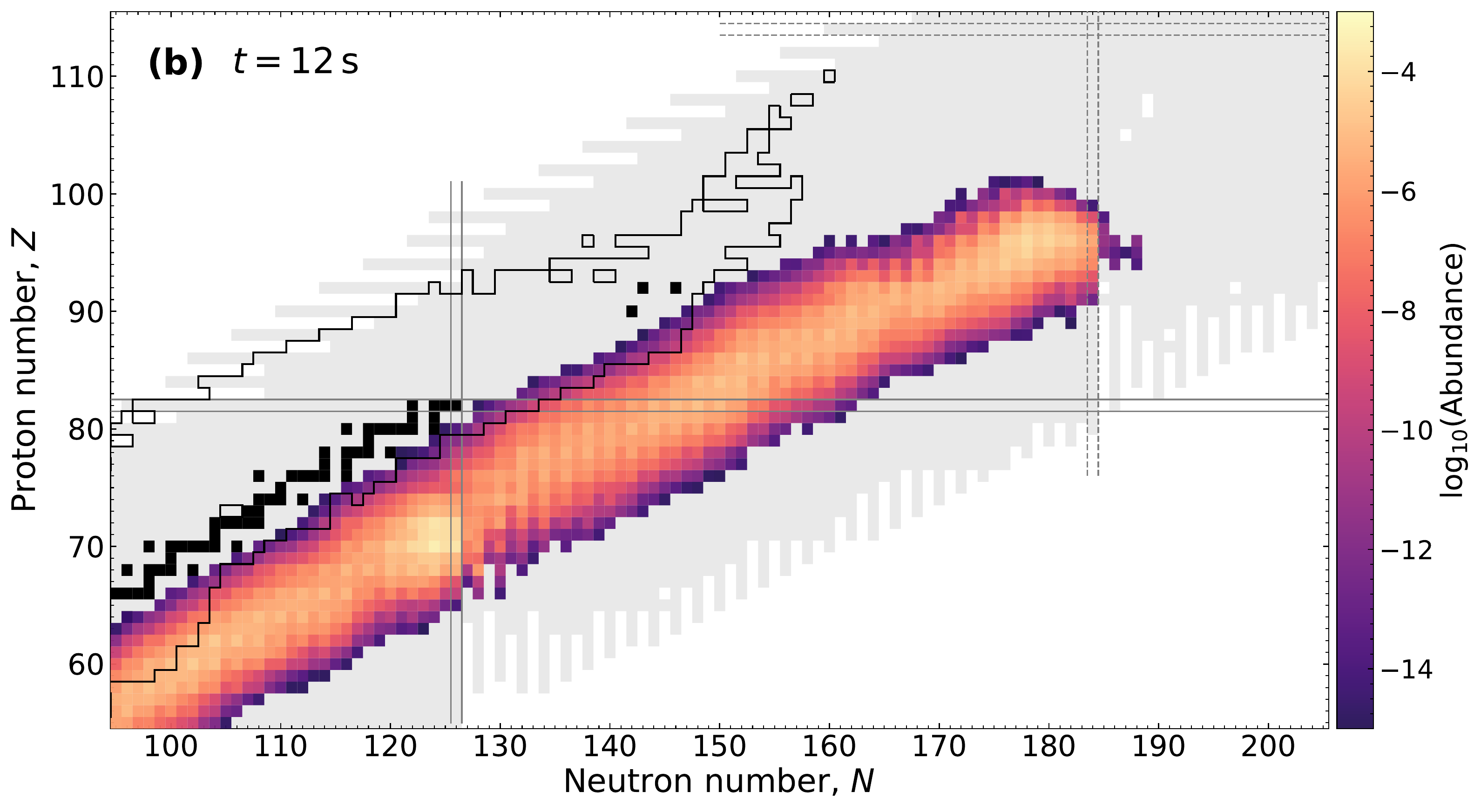}
 \includegraphics[width=140mm]{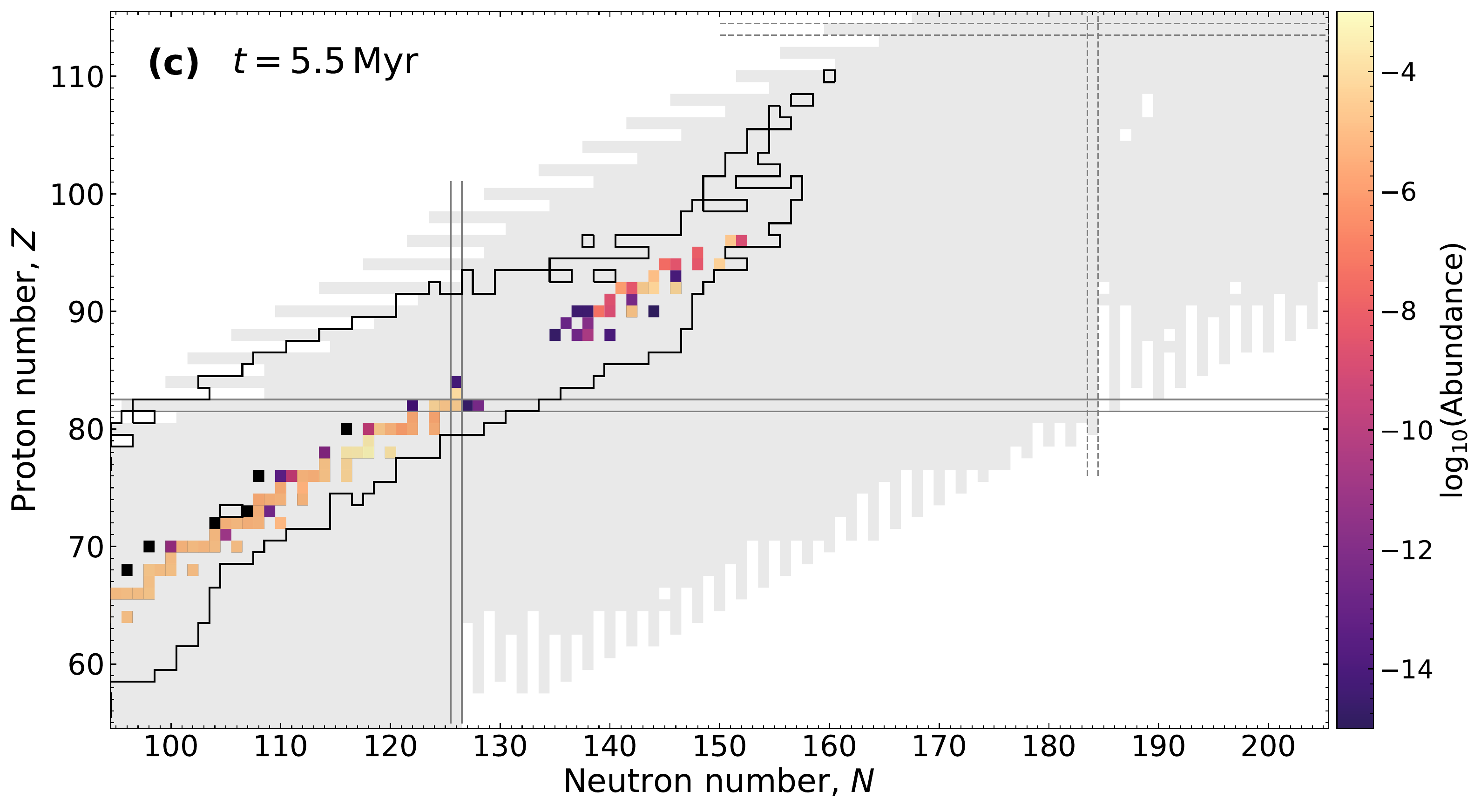}
\caption{Snapshots of nuclei produced in an \emph{r} process network calculation using astrophysical conditions from the neutron-star merger simulations of \cite{Rosswog2014} and nuclear data from \cite{Mumpower2018} for three key times: (a) before (n,$\gamma$) freezeout, (b) decay back to stability, and (c) well after the \emph{r}-process event. Colored boxes show the abundances of the nuclei at each moment in the calculation. Stable nuclei are shown as black squares, and the limit of mass measurements from AME2020 \cite{Wang2021} is shown by the black outline. The grey shaded region shows the theoretical extent of bound nuclei that may exist according to the FRDM2012 mass model \cite{Moller2012}, and the vertical and horizontal bars at $N=126$ and $Z=82$ denote closed nuclear shells.
\label{fig:snapshots}
}
\end{figure*}

After this trade-off between neutron-induced reactions and $\beta$ decay has continued for some time (generally within a second or so), the neutrons will be mostly or entirely exhausted, after which $\beta$ decay overtakes neutron capture as the dominant transmutation mechanism.
Figure \ref{fig:snapshots} shows the abundances of each nucleus at freezeout (panel `a') and other key times in an \emph{r} process calculation using neutron-rich conditions (trajectory from \cite{Rosswog2014} with nuclear physics as in \cite{Mumpower2018}).
After freezeout, the nuclei begin their long series of decays, as show in Fig.\ \ref{fig:snapshots}b, eventually terminating at the stable and extremely long-lived nuclei and beginning an isochronic phase of the \emph{r} process (Fig.\ \ref{fig:snapshots}c) \cite{Sprouse2022}.
Save for a handful of late-time $\alpha$ decays, spontaneous fission events, and latent neutron capture that can adjust the final abundances \cite{Zhu2018,Wu2019}, the abundances in Fig.\ \ref{fig:snapshots}c show the stable and long-lived nuclei created by the \emph{r} process that, e.g., are now observed in the Solar system and throughout the universe.

Both during the \emph{r} process itself as well as after freshly synthesized nuclei have decayed to their final, stable/long-lived configurations, the closed neutron shells at $N=50$; 82; 126; and, possibly, 184 manifest as relatively over-abundant `peaks' in the abundance pattern.
For example, the peaks in the Solar system isotopic pattern in Fig.\ \ref{fig:generic_solar} at $A\sim 80$, 130, and 195 are a direct consequence of the closed neutron shells.
(The sharper peaks at higher mass numbers in Fig.\ \ref{fig:generic_solar} are also from the same closed neutron shells, but occur from the \emph{slow} neutron-capture process, which populates those shells at higher charge numbers, and hence, higher mass numbers overall.) 
These peaks are explained simultaneously by an increase in half-lives along the shell closure and the sudden drop in neutron separation energies that occurs just after the shell closures are crossed \cite{Mumpower2015a}.
The net effect of which is to strongly impede the nuclear flow across such isotones, leading to the retention of nuclei at a neutron richness considerably less than that found elsewhere throughout the chart of nuclides \cite{Mumpower2015b}.
Because the $\beta$-decay rates for these closed-shell nuclei are slower than surrounding counterparts, the closed shells act to restrict the flow of nuclear abundance to heavier nuclei, and nuclei will tend to accumulate around these discrete neutron numbers.
Notice in {Fig.\ \ref{fig:snapshots}a} that nuclei at the closed neutron shell $N=126$ indeed show the largest abundances in the calculation.
After freezeout---and depending on the extent of latent neutron capture---the nuclei once populated at the closed $N=126$ shell maintain their high abundances along the same \emph{isobars} ($A\sim 195$), but the high abundance peak shifts to lower neutron numbers.
This shift of the abundance peak can just be begun to be seen in Fig.\ \ref{fig:snapshots}b by the slight shift to lower neutron numbers of the high-abundance region at $Z=70$.
Finally, at late times this high-abundance feature (produced pre- and mid-freezeout as an effect of the neutron shell closure at $N=126$) now populates $N\sim 118$, as can be seen in Fig.\ \ref{fig:snapshots}c.

The proposed closed neutron shell at $N=184$ is particularly relevant for the superheavy elements, as it is around the $N=184$ and $Z=114$ doubly magic region that the Island of Stability is speculated to occur \cite{Chapman2020}.
We discuss next the region beyond $N=126$, where the abundances of nuclei during a standard \emph{r} process depend on the complicated interplay of reactions beyond just neutron capture and $\beta$ decay.

\subsection{Production of the heaviest elements}
\label{sec:production}

As described in the previous section, a basic canonical picture of the \emph{r} process involves neutron capture, its inverse process of photodissociation, and $\beta$ decay.
However, as soon as an \emph{r} process produces the heaviest elements, particularly those well beyond Pb, the situation becomes substantially more complicated, and the effects of a more diverse set of processes---such as $\alpha$ decay and fission---may significantly contribute to the overall dynamics of the ensuing nucleosynthesis.
The competition between inflowing and outflowing material at high charge and mass ultimately determines abundance of heavy elements produced by the \emph{r} process, including, possibly, the superheavies.

Recalling Fig.\ \ref{fig:rpath}, neutron capture and $\beta$ decay bring material to the actinide region.
Once occupying the actinides, nuclei have a chance to undergo fission through neutron-induced, $\beta$-delayed, and/or spontaneous fission channels \cite{Mumpower2018, Hao2022}.
As described in Sec.\ \ref{sec:fission}, nuclei with proton number larger than $Z\sim90$ may be driven to fission by an incoming neutron \cite{Panov2003,Martinez-Pinedo2007}.
Similarly, fission can follow the $\beta$ decay of a nucleus if that decay places it in a sufficiently excited state \cite{Thielemann1983,Ghys2015,Mumpower2018}.
For some theoretical models, $\beta$-delayed and neutron-induced fission can ultimately terminate the \emph{r} process and prevent very high-$A$ nuclei from being synthesized.
This effect is apparent in Fig.\ \ref{fig:barrier} by the extremely low fission barrier heights near $Z=100$ and just beyond $N=184$.

\begin{figure*}
 \centering
 \includegraphics[width=0.85\textwidth]{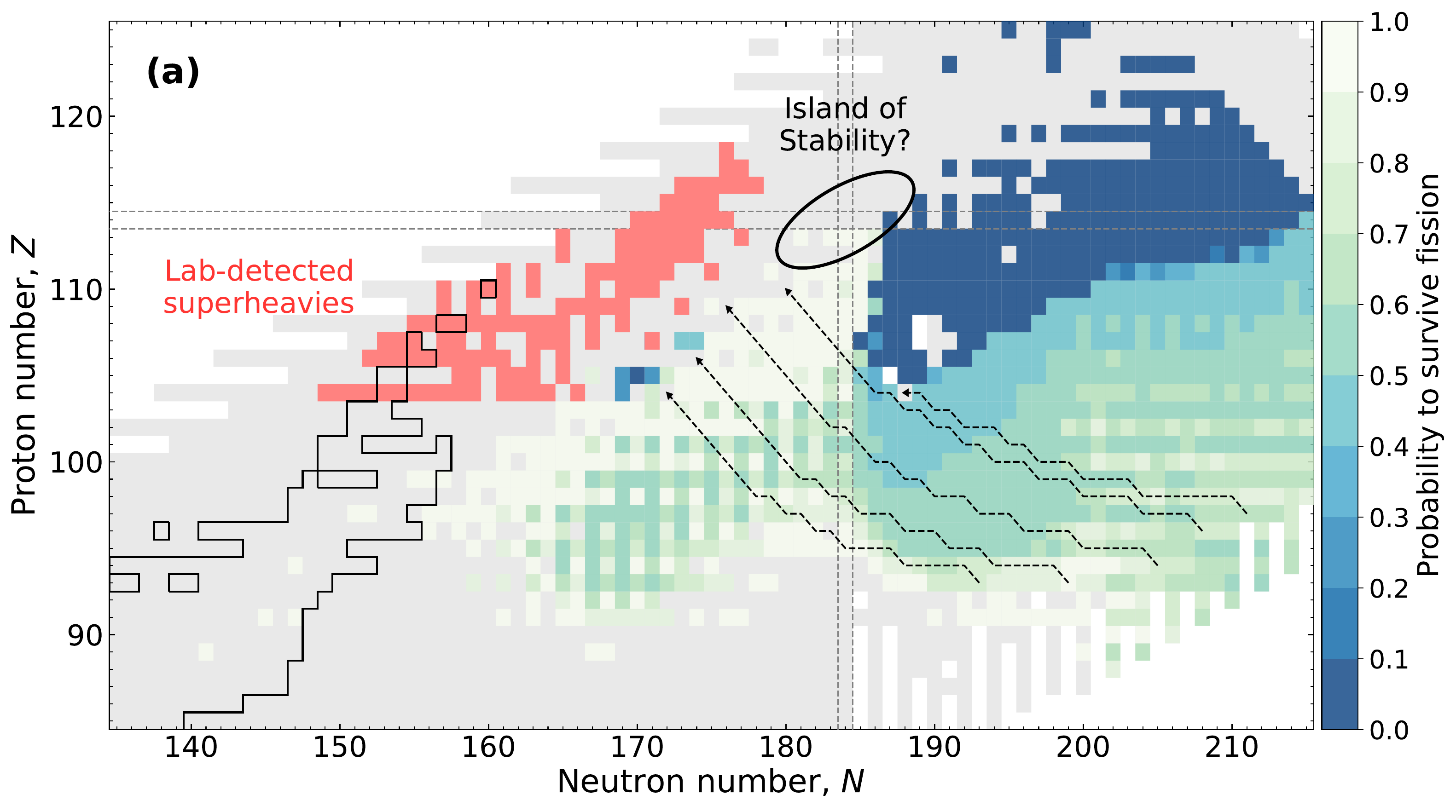}
 \includegraphics[width=0.85\textwidth]{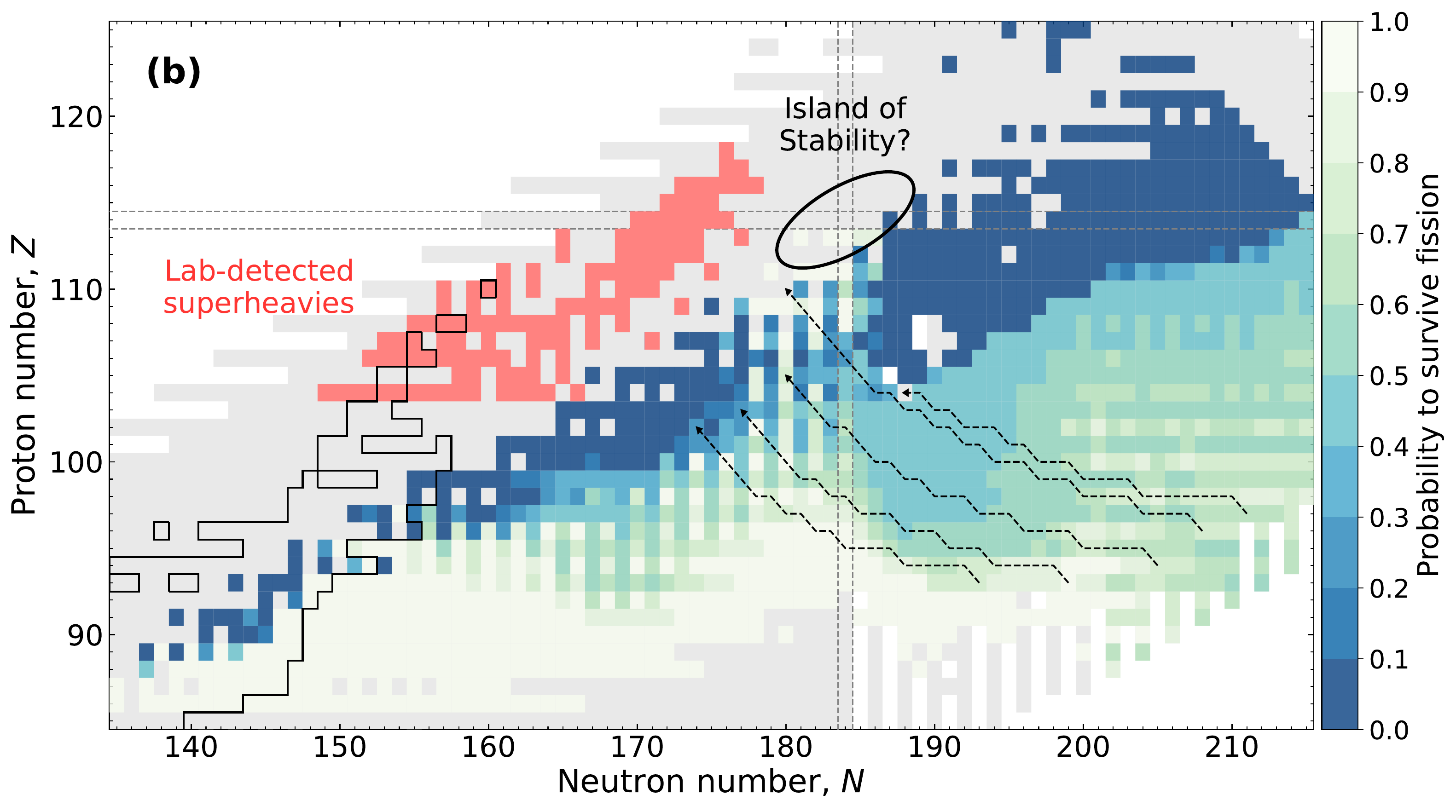}
\caption{Most likely $\beta$-decay paths (dashed) of some neutron-rich nuclei that can populate the superheavy elements as they traverse through regions where there is a probability to fission (colors), calculated using decay and reaction data based on the FRDM2012 mass model \cite{Moller2012,Moller2016} and FRLDM fission barrier heights \cite{Moller2015}. Only $\beta$-delayed fission and spontaneous fission are included in panel (a), and panel (b) additionally includes the neutron-induced fission channel. The black-outlined region shows the extent of current nuclear mass measurements from the AME2020 compilation \cite{Wang2021}, and the grey boxes show a theoretical extent for what nuclei could possibly exist. Superheavy elements that have been observed in laboratory settings are marked with red, and the black circle indicates the region of the theoretical superheavy `Island of Stability.'
\label{fig:decay_fission}
}
\end{figure*}

Modern simulations of the \emph{r} process suggest a final fission termination point between $A\sim275$ and $A\sim300$ \cite{Eichler2015,Mumpower2018,Goriely2013}.
The nuclei populated in this region of the chart of nuclides are certainly superheavy in the sense that they have an extreme neutron excess.
Whether superheavy \textit{elements} are created this far from stability early on in the \emph{r} process as well as their possible survival to more stable species remains an open question.
These questions are related to the maximum termination point of the \emph{r} process, which is also uncertain.
Sufficiently high fission barriers for these extremely neutron-rich species would raise the terminus of the \emph{r} process to even higher mass numbers \cite{Shibagaki2016}.

Once the neutron-flux drops and the $\beta$-decay timescale dominates over neutron-capture, the newly synthesized neutron-rich nuclei can decay to even heavier elements, whereby they have another chance to populate the superheavies.
Figure \ref{fig:decay_fission} shows paths along which some neutron-rich nuclei that could be made in an \emph{r} process may $\beta$ decay.
The most neutron-rich nuclei are more likely to emit at least one neutron after $\beta$ decay ($\beta$-delayed neutron-emission), which further decreases the mass of the nuclei that are ultimately formed at the end of those $\beta$-decay chains.
However, these nuclei are not completely free to $\beta$-decay undisturbed; during these $\beta$-decays, fission processes can continue to evacuate material from the heavy-element region.
Sufficiently low fission barrier heights in this region can cause nuclei to prefer fission over $\beta$ decay and therefore throttle production of the actinides and superheavies.

The green and blue colors in Fig.\ \ref{fig:decay_fission}a show the probability that a nucleus will survive fission and $\beta$ decay in the absence of neutron-induced fission.
These probabilities were calculated by taking the ratio of the $\beta$-decay rate to the sum of the $\beta$-decay, $\beta$-delayed, and spontaneous fission rates (from the FRDM2012 mass model \cite{Moller2012,Moller2016} and FRLDM fission barrier heights \cite{Moller2015}) on a per-nucleus basis.
By taking the product of the probabilities along a particular $\beta$-decay path (e.g., one of the black dotted paths), we can obtain an upper limit for how likely it is that a particular nucleus at freezeout will transmute into a superheavy species.
To create, e.g, $^{290}$Ds ($Z=110$), a progenitor $^{304}$Cm ($Z=96$) nucleus would have to traverse through several regions where the probability to fission exceeds the probability to $\beta$ decay.
In total, the cumulative probability that $^{304}$Cm and the daughter nuclei that are created by subsequent $\beta$ decay will survive each opportunity to fission and create $^{290}$Ds is found to be about 0.1\% in the absence of neutron-induced fission.

When the neutron-induced fission channel is included (i.e., by adding an additional term to the rate ratio), even at low temperatures and neutron fluxes, neutron-induced fission at late times when the nuclei are closer to stability can thwart superheavy-element production, as shown in Fig.\ \ref{fig:decay_fission}b.
In the case of $^{290}$Ds, even a low neutron flux decreases the total probability of production by more than an order of magnitude.
On the lower-mass side, there is a small window with lower probabilities to fission where nuclei can decay into the superheavy series.
For example, in the absence of neutron-induced fission $^{294}$Pu ($Z=94$) has about a 0.8\% chance to survive every fission event along its most-probable decay chain and create $^{283}$Hs ($Z=108$).
(Neutron-induced fission changes this endpoint nucleus to $^{283}$Db ($Z=105$) with a combined survival probability of 0.04\%.) 
In the next section, we will explore how these predictions vary for different assumptions of theoretical nuclear physics properties.

\subsection{Sensitivity to nuclear physics}

In the extremely neutron-rich regions of the chart of nuclides, all properties of nuclei entering into \emph{r} process simulations depend on theory calculations, and predictions by theory may vary.
For example, the strength of the closed neutron shells affects how long nuclei remain at that nucleus, eventually determining how much mass occupies the closed shells.
Neutron shells that are too strong effectively stockpile abundance at the closed shells, which may prevent material from reaching the actinide region and have important consequences for superheavy elements production \cite{Schatz2002}.
In particular, the $N=126$ is the highest known spherical neutron shell closure.
Though theories predict the next spherical neutron shell closure at $N=184$, its existence has not yet been confirmed.
In modern, dynamic simulations, neutron-rich \emph{r} processes that proceed beyond the $N=126$ shell closure are typically terminated around $N=184$ by fission processes, as discussed in Sec.\ \ref{sec:fission} and \ref{sec:production}.
The strength of this closed shell plays an additional role in the production of heavy and superheavy nuclei moderating late-time neutrons.
Recent microscopic fission models \cite{Giuliani2020} show that material accumulated around closed shells like $N=184$---resistant to neutron capture---can allow more late-time free neutrons, which may in turn induce fission on the heavy nuclei decaying to stability, for example as is the case in Fig.\ \ref{fig:decay_fission}b.
On the other hand, if the $N=184$ shell closure does not exist, or is weakened as a function of neutron excess, it is possible that the \emph{r} process terminates at higher mass numbers, providing an easier pathway for the population of superheavy nuclei.
Such considerations are sensitive to the behavior of nuclear reactions at large neutron excess \cite{Patra2009}.

Similarly, $\beta$ decay determines how quickly material traverses between (and beyond) the closed neutron shells.
This is particularly important for the $N=126$ closed shell as it acts as a gatekeeper to the production of the heaviest elements \cite{Tang2020}.
Decay rates that include First-Forbidden transitions for nuclei above the $N=126$ shell closure increase the $\beta$-decay rates by over an order of magnitude on average compared to model predictions that do not fully include these types of transitions (e.g., \cite{Marketin2016} versus \cite{Moller2019}).
As a consequence, heavy nuclei spend very little time in the $N>126$ region before reaching the fission region.
In the end, overall less material populates the actinide region by the time $\beta$ freezeout occurs, resulting in a decreased likelihood that the superheavy elements will be robustly populated.
Other studies of $\beta$-decay involve experimentally characterizing $\beta$-delayed neutron emission for neutron-rich nuclei \cite{Dimitriou2021,Hall2021}.
Not only can late-time neutron emission increase the probability of heavy nuclei to fission, but understanding the competition the between $\beta$-delayed neutron-emission and $\beta$-delayed fission channels will also inform how much material is mapped to lower mass numbers in the \emph{r} process due to $\beta$-delayed fission.

If the theoretical (and experimental) mass and $\beta$-decay predictions are generous enough to abundantly populate the actinide region during the \emph{r} process, the heavy nuclei will then have to contend with fission, another source of uncertainty in nuclear theory.
As shown in Fig.\ \ref{fig:barrier}, there are significant regions with low barriers to fission, which is intrinsically related to the fission rate.
Such low barrier heights cause fission to be the dominant transmutation channel, decreasing the probability that a heavy nucleus will survive as it undergoes $\beta$ decay though the low fission barrier region \cite{Petermann2012,Goriely2015b}.
Low fission barrier heights beyond $N=184$ inhibit the production of superheavy elements, particularly as nuclei pass through a region where the probability to undergo $\beta$-delayed fission is nearly 100\% \cite{Mumpower2018}.
Fission barrier heights---and therefore the probability to undergo fission---differ between theory predictions.
Theories that predict higher fission barriers and a weaker (or no) $N=184$ shell closure have an increased probability of populating heavier masses that may decay to superheavy elements, provided that the barrier heights are sufficiently high that $\beta$-delayed fission is also suppressed.

Another interesting possibility that would impact the population of actinides and superheavies is the emergence of an alternative shell closure in this region of superheavy nuclei.
A deformed or spherical shell closure below $N=184$ would hold material up at lower mass number, likely leading to a relatively larger population of the lightest actinides.
Conversely, the absence or weakening of the proposed $N=184$ closure paired with another higher closure could lead to a larger production of the heaviest acitnides or superheavy elements, assuming fission does not substantially interfere during the decay back to stability.

\begin{figure}
 \centering
 \includegraphics[width=\columnwidth]{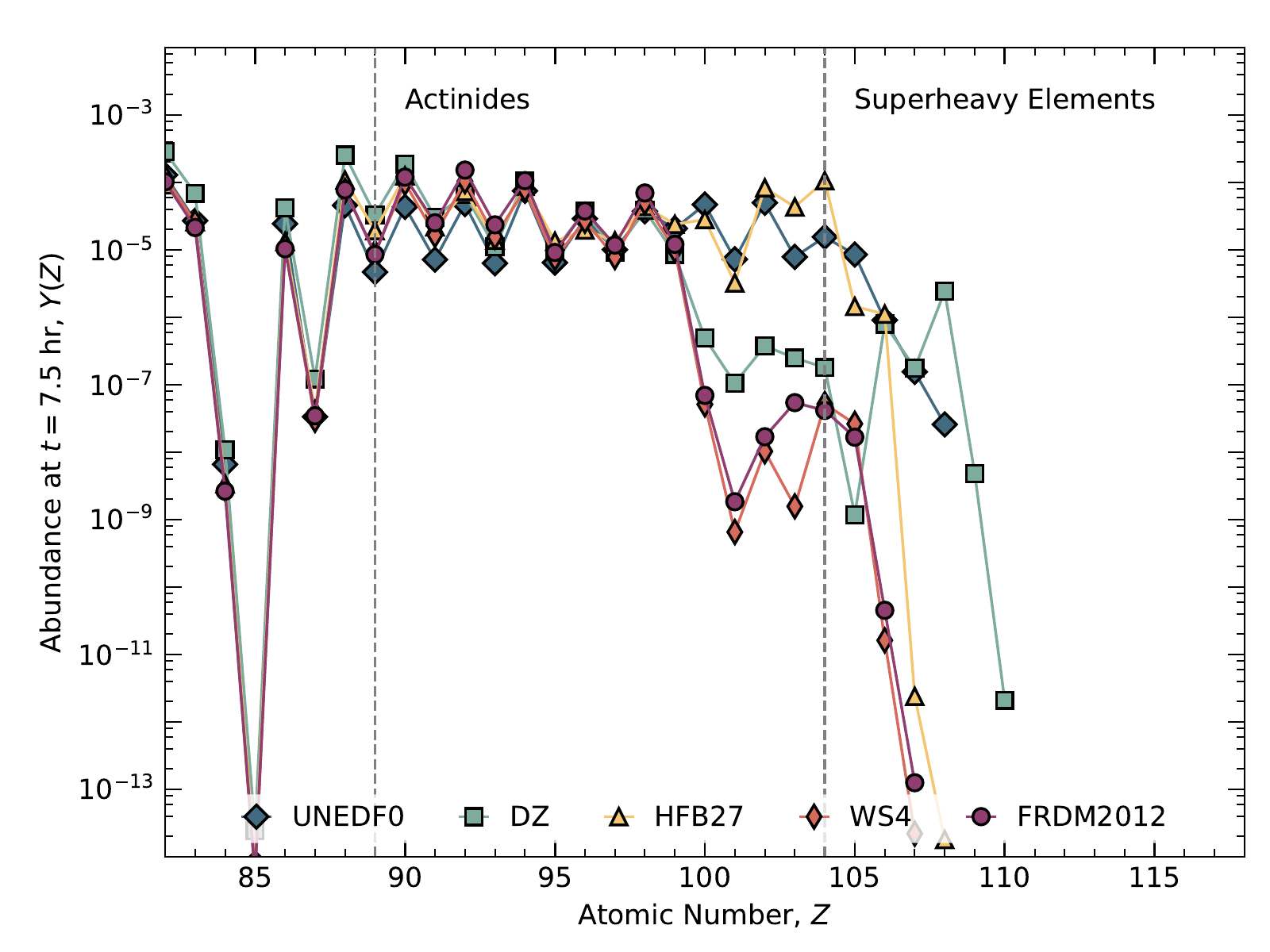}
 \caption{Total elemental abundances at about seven hours in an \emph{r}-process calculation using different nuclear mass models and fission barrier heights. For the UNEDF0 and HFB27 models, barrier heights from HFB14 were used. Both WS4 and FRDM2012 use FRLDM fission barrier heights \cite{Moller2015}, and DZ uses KTUY barriers \cite{Koura2014}.
\label{fig:Zchains}}
\end{figure}

Figure \ref{fig:Zchains} shows how differing mass model and fission barrier height combinations can produce different elemental abundances in an \emph{r}-process simulation.
To explore the current variation in model predictions, HFB14 barriers were used in combination with UNEDF0 and HFB27 masses, FRLDM barriers were used for FRDM2012 and WS4 masses and KTUY barriers were used for DZ masses.
At approximately seven hours after an \emph{r}-process event, a relatively robust pattern emerges for elements below $Z\sim98$, whereas above Cf, the predictions vary substantially.
The heaviest elements are produced with the DZ/KTUY dataset, resulting in a termination around $Z\sim108$.
The HFB model produces the highest yield of superheavies while the masses with FRLDM barriers represents a more pessimistic case.
According to their primarily $\beta$-decay population mechanism, the superheavies that would be synthesized by the \emph{r} process would naturally be on the neutron-rich side of the chart of nuclides and hence likely to be more neutron-rich than what has currently been created in the laboratory.

The select models used here suggest that thorium ($Z=90$), uranium ($Z=92$), and plutonium ($Z=94$) are among the most populated actinide species around 7.5 hours after an \emph{r}-process event.
Subsequent $\alpha$-decay from heavier actinides can continue to to contribute to their production on much longer timescales than that of the neutron capture phase of the \emph{r} process.
Above californium ($Z=98$) the models begin to substantially diverge in their production of the heaviest elements.
The KTUY fission barrier model, with its high fission barriers, consistently affords the ability to produce the nuclei with highest proton number, even with variation in the assumed masses (not shown).
It is also found that HFB models consistently produce the highest population of superheavies; in the case of Fig.\ \ref{fig:Zchains}, the bulk of the population resides in rutherfordium ($Z=104$).
The diversity of species populated in Fig.\ \ref{fig:Zchains} will result in a significant variation in the final decay pathways of the heaviest elements, which in turn influences the potential for observations that occur on longer timescales.

Revisiting the values in Fig.\ \ref{fig:decay_fission} with a more optimistic choice like HFB27, the cumulative probability that $^{304}$Cm survives fission to populate $^{290}$Ds is about 7.6\% without neutron-induced fission, and 1.8\% with it active.
For a slightly lower-$Z$ superheavy, say $^{283}$Hs, the decay path from ${299}$Pu goes between regions of low barrier heights, leading to a cumulative survival probability of 63\%.
Masses and fission barrier heights thus have a significant effect on the superheavy elements that can be produced by the \emph{r} process.

It is important to recognize several points here regarding the simulations shown in Fig.\ \ref{fig:Zchains}; even in optimistic scenarios (e.g.\ DZ/KTUY) with significant superheavy elements created, they are likely to be found in trace amounts relative to the abundances of lighter nuclei which can be populated in many more, less extreme, astrophysical conditions.
Second, simulations presented here do not fully estimate all the nuclear uncertainties involved in the production of the heaviest nuclei.
Additional uncertainties, particularly with $\beta$ half-lives will serve to increase the variations observed in Fig.\ \ref{fig:Zchains}.

Given the orders of magnitude variation found in the production of superheavy elements with current modeling, extensive work involving measurements and theoretical calculations are required to resolve this situation.
Continued efforts to understand the fission of heavy neutron-rich nuclei will clarify the likelihood of heavy nuclei to survive their long $\beta$-decay chains and eventually populate the superheavy elements \cite{Erler2012,Ishizuka2022}.
Observational implications of fission will be discussed in Sec.\ \ref{sec:observation}.

\subsection{Sensitivity to astrophysics}

Not only is the production of superheavy elements sensitive to the as-of-yet unknown nuclear physics of neutron-rich nuclei, but it is also sensitive to the astrophysical conditions responsible for an \emph{r}-process event.
Temperature, entropy, density, and neutron-richness shape the initial elemental composition---which nuclei are present when the \emph{r} process begins---as well as the reaction rates between nuclei during the \emph{r} process, in part determining how far in mass number the \emph{r} process progresses.
For example, at the same temperature and density, astrophysical conditions with a higher neutron-richness (more free neutrons available for capture) will generally produce a stronger \emph{r}-process, and thus a higher probability of producing the heaviest elements.
The dependence of isotopic production on \emph{r}-process dynamics is not unique to the discussion of superheavy elements.
For example, the formation of the rare-earth nuclei has been shown to depend on the conditions of \emph{r} process freezeout  \cite{Surman1997,Otsuki2010,Arcones2011,Goriely2013,Mumpower2016a,Vassh2019}, and the production of the heaviest elements has also been thoroughly explored in previous work (e.g., \cite{Kratz1993,Qian2007,Eichler2016,Eichler2019}).

\begin{figure}[t]
 \includegraphics[width=\columnwidth]{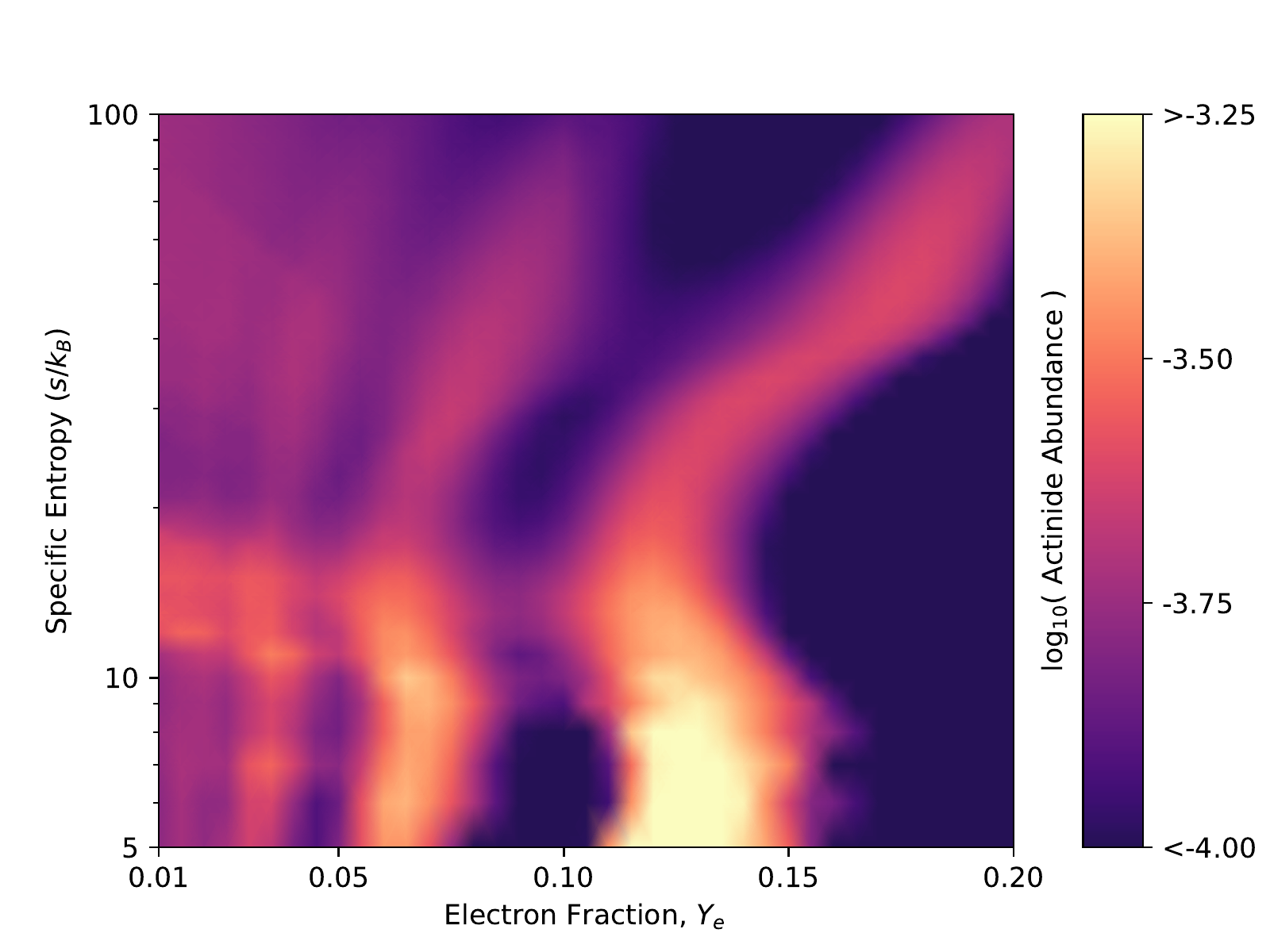}
\caption{Actinide production as a function of entropy (vertical axis) and electron fraction (horizontal axis) for parameterized astrophysical conditions with fixed expansion timescale $\tau=10\ \textrm{ms}.$} 
\label{fig:yact_vs_conditions}
\end{figure}

Figure \ref{fig:yact_vs_conditions} highlights an example of the dependence of heavy-element production on two astrophysical parameters in a standard \emph{r} process: specific entropy ($s$) and initial composition ($Y_e$).
The trajectories used for the construction of Fig.~\ref{fig:yact_vs_conditions} derive from the parameterization used in \cite{Lippuner2015}, where we have fixed the timescale parameter $\tau = 10\ \textrm{ms}$.
Generally, lower $Y_e$---and thus more neutron-rich conditions---yield a higher actinide abundance which is indicated by bright yellow.
Similarly, a higher entropy generally enables even moderately low-$Y_e$ conditions to produce significant actinide material since high-entropy conditions allow more photodissociation of neutrons from nuclei.
Noting the trends in Fig.\ \ref{fig:yact_vs_conditions}, the production of the heaviest elements is clearly not straightforwardly quantifiable as a simple function of astrophysical parameters.
In any case, the greatest chance of observing the superheavy elements would likely result from an astrophysical environment in which the abundance of the heavy elements is maximized.

Currently, the conditions that favor heavy and superheavy element production are low-entropy, low-$Y_e$ astrophysical sites, generally agreeing with early studies of superheavy element production in \emph{r} processes that use parameterized calculations \cite{Ohnishi1972,Schramm1973,Berlovich1974,Mathews1976}.
Modern simulations of binary neutron star mergers robustly predict low-entropy, low-$Y_e$ conditions \cite{Goriely2011,Korobkin2012}, which could prove suitable for superheavy-element production.
Neutron star mergers contain several interesting sites possible for \emph{r}-process nucleosynthesis including a post-merger accretion disk \cite{Surman2006,Surman2008} and tidally stripped ejecta \cite{Just2015}.
Complex simulations with Monte-Carlo radiation transport of the accretion disk \cite{Miller2019,Foucart2020} currently do not produce significant quantities of the heaviest elements, but this conclusion depends strongly on uncertainties in neutrino physics which controls the morphology of the disk \cite{McLaughlin1996,Surman2005,Patton2012,Zhu2016}.
Tidal ejecta may be a more favorable location for the production of the heaviest elements owing to generally lower initial electron fractions and less time for neutrinos to interact with the material \cite{Radice2016,Bovard2017,Radice2018a,Radice2018b,Hotokezaka2015}.
Complete simulations from the inspiral phase until creation of the post-merger disk will be very valuable in assessing the capacity of heaviest element synthesis in neutron star mergers.

Other compact object mergers of interest include black-hole neutron-star (BHNS) mergers which also have the potential to produce the heaviest elements \cite{Nishimura2016,Roberts2016}.
The study of BHNS binaries offers for exploration a wide range of physical processes beyond nucleosynthesis, including the neutron star equation of state, stellar evolution, high-energy astrophysics, and the expansion of the universe \cite{Foucart2020b}.
However, black-hole neutron-star mergers may be unlikely multi-messenger sources due to the requirement of high-spin for an electromagnetic signal which is further compounded by limitations of current observing technologies \cite{Fragione2021}.

Finally, it remains possible for magnetorotational supernovae or collapsars to reach suitable conditions for the formation of the heaviest elements \cite{MacFadyen1999,MacFadyen2001,Surman2011,Nakamura2015,Siegel2019,Miller2020}.
Further investigation is warranted on this front, particularly with the starting configurations of the disk which strongly influences the resultant nucleosynthesis.

\section{Observation of the heaviest elements}\label{sec:observation}

We have discussed the properties of the heaviest nuclei from a nuclear perspective and how those properties play into the potential astrophysical production of these elements.
If superheavy elements are made in the universe, where and how can we observe them, if at all? 
The definition of `observation' in this context extends from direct measurements through, e.g., atomic mass spectrometry and stellar spectroscopy, to indirect ones through, e.g., heavy-element decay products and lightcurves following explosive astrophysical events.
Here we review a brief history of translead observations both in the Solar system and outside of it and discuss potential observations regarding the heaviest elements that may be formed in the \emph{r} process.

\subsection{Actinides in the Solar system}

The existence of heavy elements on Earth and in meteorites gives some indication of the astrophysical events that produced those elements before Solar system formation.
Studies that seek to explain the last enrichment event by the \emph{r} process before the Solar system formed frequently employ not only the heavy elements, but also their short-lived radioactive decay products \cite{Lugaro2018,Wallner2021,WangX2021}.
Understanding the production of heavy elements and their decay products can constrain galactic chemical evolution studies by elucidating which \emph{r}-process sites dominate at which times in galactic history \cite{Wanajo2006,Wehmeyer2015,Cote2018,Ojima2018}.
In any case, the measurements of translead nuclei created before Solar system formation is constrained by the half-lives of those elements.
Recall from Fig.\ \ref{fig:t12heavy} that all known translead nuclei are unstable, and the half-lives of superheavy elements generally become shorter with increasing atomic number.
Even if superheavy elements were present at Solar system formation, they likely would have decayed to less than trace abundances over the 4.6-Gyr age of the Solar system.

Starting in the early 1960's, fossil tracks were found in naturally occurring mica, believed to be made from the fragment from the spontaneous fission of $^{238}$U \cite{Price1962}.
These observations followed a study on artificially irradiated mica that discovered uranium fission fragments can leave such a record \cite{Silk1959}.
Extending this logic, it was believed that fossil tracks from cosmic rays could be found in cosmically irradiated meteorites, and such tracks were identified for medium nuclei and heavy nuclei, including $^{244}$Pu, within the decade \cite{Maurette1964,Fleischer1965,Fleischer1967,Fleischer1967b}.
Extrapolating further still, experimental undertaking were designed in the specific search for superheavy nuclei in meteorites \cite{Ginzburg2005,Aleksandrov2008}.
Results from these experiments claim detections from superheavy nuclei as well as lighter, transuranic species that could possibly be decay products from the same long-lived superheavies
\cite{Bagulya2013,Alexeev2016,Alexandrov2019,Alexandrov2022}.

\begin{figure*}
 \centering
 \includegraphics[width=0.42\textwidth]{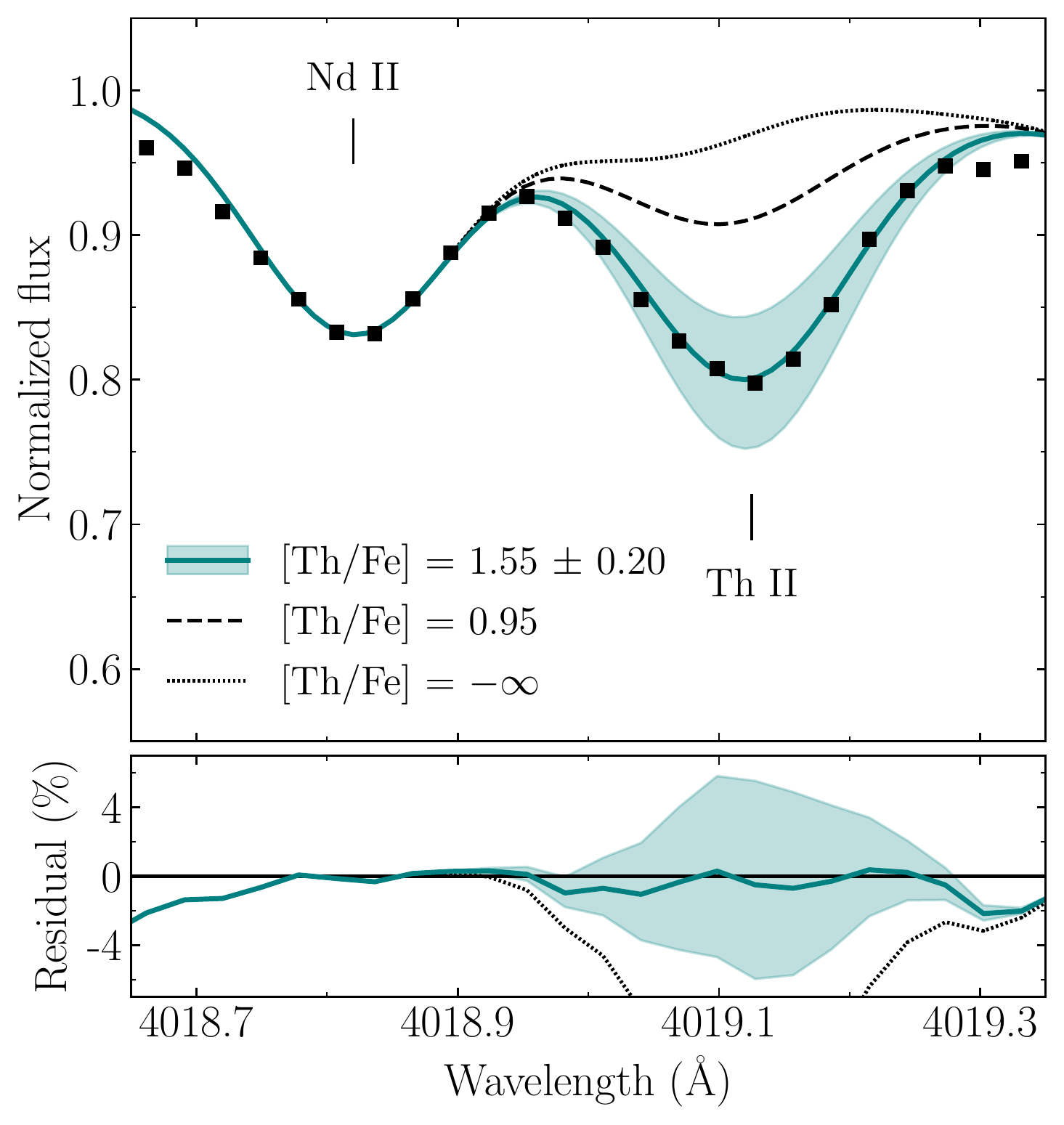}\qquad
 \includegraphics[width=0.42\textwidth]{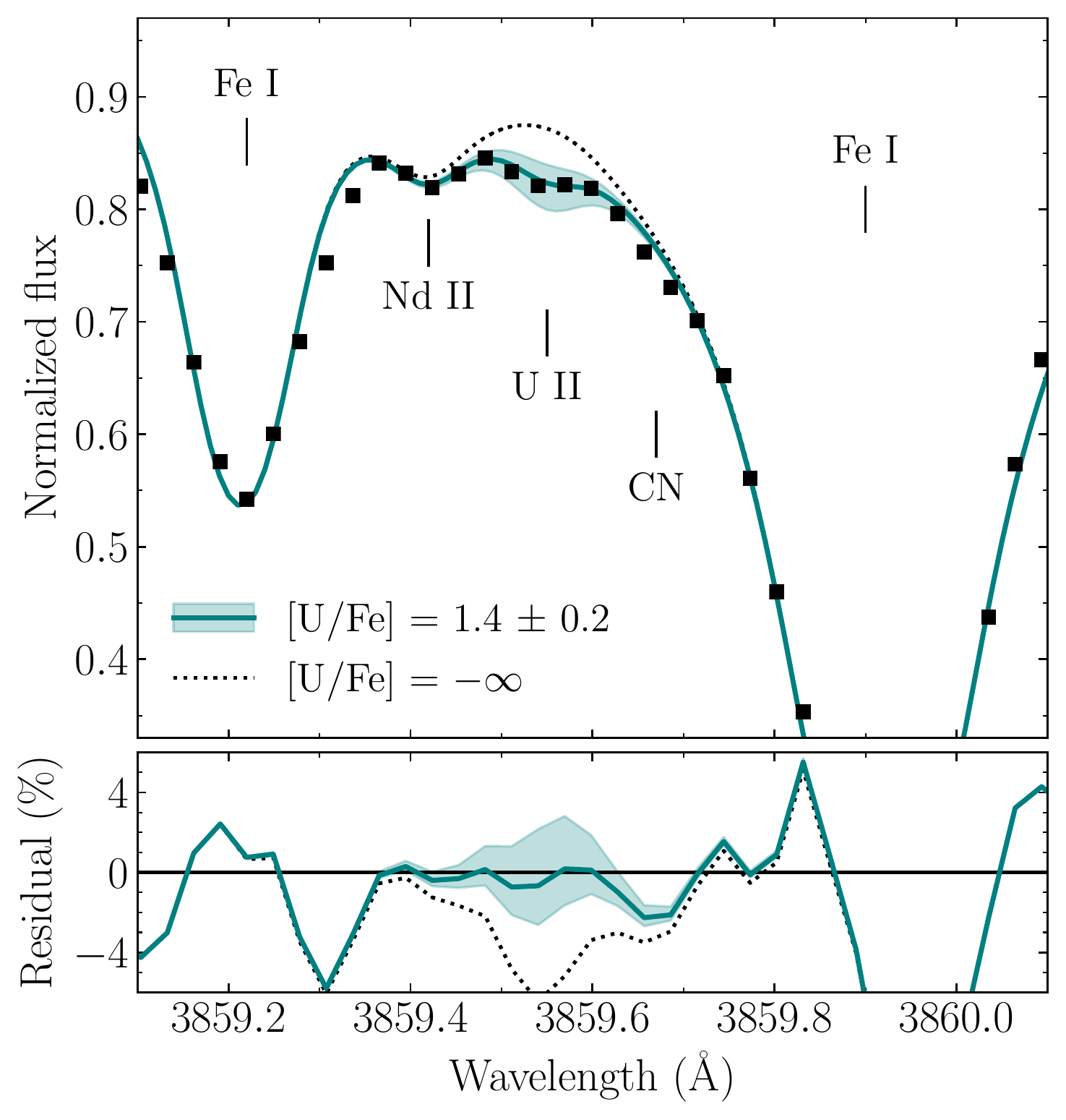}
 \caption{Observed stellar spectrum (points) in wavelengths centered on the absorption features of Th (left) and U (right) compared to models (colours) used to derive the abundance of the element in the star. Dotted lines show model spectra with no Th and U present. Absorption features of notable elements and molecules are also labelled.}
\label{fig:Th_spectrum}
\end{figure*}

Inferring the existence of short-lived mother nuclei by measuring the abundances of their longer-lived daughter nuclei has been applied to other actinides elements as well.
For example, after a series of $\beta$ and $\alpha$ decays, $^{247}$Cm ($T_{1/2}=15.6$~Myr) eventually transmutes into the much longer-lived $^{235}$U ($T_{1/2}=708$~Myr) \cite{Stirling2005}.
The ratios of long-lived actinide species like $^{235}$U, $^{238}$U ($T_{1/2}=4.47$~Gyr), and $^{232}$Th ($T_{1/2}=14$~Gyr) are often used to calculate the time between the \emph{r}-process event that produced them and the time they are observed (i.e., present day) through a method known as `cosmochronometry' \cite{Dicke1969,Truran1998,Cowan1999,Goriely1999,Schatz2002,Wanajo2002,Otsuki2003,Brennecka2010}.
In the Solar system, ages calculated from the ratio of $^{235}$U to $^{238}$U have shown some disagreement from those calculated from other ratios, e.g., $^{232}$Th to $^{238}$U.
Since all actinides are necessarily \emph{r}-process-only nuclei, they must all have been made in the same event(s), and variations between the ages inferred from the ratio point to an inconsistency in understanding their production rather than variations in the age of the material itself.
To solve this inconsistency, it has been suggested that the decay of $^{247}$Cm, which would have been present in the early Solar system, causes a residual enhancement to $^{235}$U \cite{Tatsumoto1980}, thus explaining the different in ages calculated with different actinide ratios.
Indeed, $^{247}$Cm found in meteorites is consistent with a single \emph{r}-process event that is responsible for the bulk of actinide material currently found in the Solar system \cite{Francois2016}.
A similar conclusion is drawn from comparing the abundances of actinides to lighter elements, i.e., through the $^{129}$I/$^{247}$Cm ratio \cite{Cote2021}.
Further observational evidence of a consistency in the $^{129}$I/$^{247}$Cm ratio would strongly suggest that both lighter \emph{r}-process nuclei and the heaviest \emph{r}-process nuclei are all created in a single event.
This ratio is also able to estimate that the time between the last \emph{r}-process event that enriched the pre-Solar gas and condensation of that gas into solids was about 100--200 Myr \cite{Cote2021}.
Observing \emph{r}-process-only nuclei like the actinides in the Solar system thus provides powerful ways of understanding their astrophysical production.

\subsection{Actinides in stars}

As mentioned in Sec.\ \ref{sec:introduction}, atomic absorption features in stellar spectra are correlated with elements present in the star's photosphere.
Through the identification and measurement of these absorption features, the elemental composition may be characterized and quantified in stars.
For example, Fig.\ \ref{fig:Th_spectrum} shows absorption features identified in a stellar spectrum over two narrow wavelength regions.
The continuum of light is normalized to unity, and significant deviations from unity correspond to atomic transitions of elements present in the stellar photosphere.
Of measured translead nuclei, only three actinide nuclei---and no superheavies---have half-lives sufficiently long for the species to be observable today in starlight: $^{232}$Th, $^{235}$U, and $^{238}$U.
The most prominent atomic transitions of Th and U in optical wavelengths are shown in Fig.\ \ref{fig:Th_spectrum}.

It is important to note that observations through stellar spectroscopy are in most cases naturally limited to deriving elemental---not isotopic---abundances.
Unless the hyperfine structure of a nucleus alters the absorption feature within measurable limits (e.g., the detector's resolving power), the shape and central wavelengths of atomic absorption features are degenerate with isotopic mass.
Only in rare cases (or extreme conditions) can isotopic abundances be inferred from stellar spectra \cite{Cowley1989,Mashonkina1999,Roederer2008}.
In the case of the actinides, it is safe to assume that the only isotope of Th present in stars is $^{232}$Th, since the half-life of the next longest-lived Th nucleus is less than 10,000 years and would have decayed away possibly even before the star was born.
Similarly, for stars sufficiently old, $^{238}$U dominates over the faster-decaying $^{235}$U, and the abundance of U derived from stars can be almost entirely attributed to $^{238}$U.
As mentioned in the previous section, this assumption does not hold for young stars like the Sun.
Beyond $^{238}$U, no heavier element has been identified in stellar spectra, and there are currently only five stars for which even the abundance of U has been derived \cite{Cowan2002,Honda2004,Frebel2007,Placco2017,Holmbeck2018}.

The identification of actinides in stellar spectra (primarily through Th) ranges over cosmic time and is not limited to the Sun or our Solar system.
Thorium has also been identified in stars similar to the Sun and could play a role in planet formation and habitability around Sun-like stars \cite{Botelho2019,Nimmo2020}.
In these `Solar twins,' the abundance of Th roughly agrees with the Sun.
Since the actinides are necessarily produced by \emph{r} process conditions, the consistency of finding Th across several stars could be considered evidence that either (1) one \emph{r} process site dominates Th production, or (2) the \emph{r} process behaves universally.

\begin{figure*}
 \centering
 \includegraphics[width=0.9\textwidth]{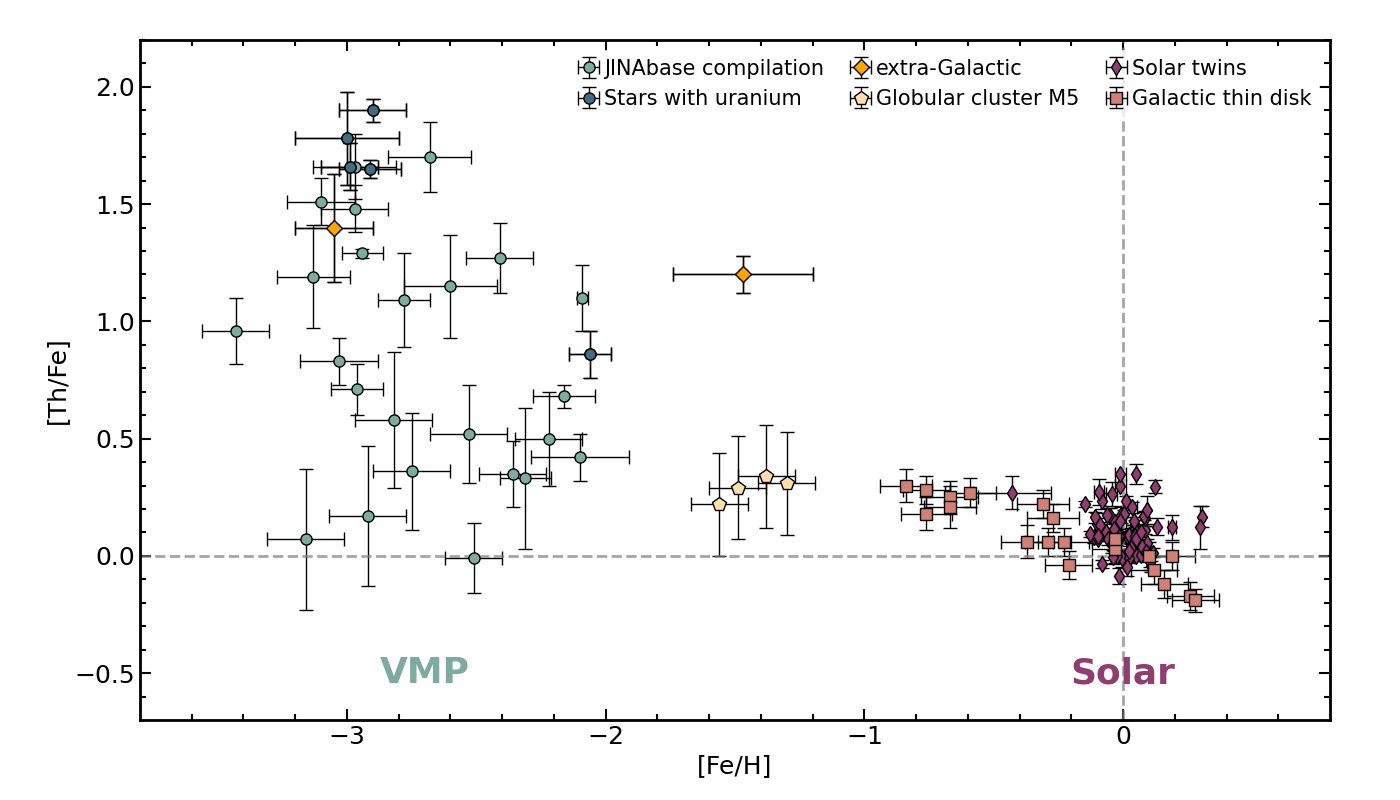}
 \caption{Thorium abundances ([Th/Fe]) for stars across a variety of metallicities ([Fe/H]) and stellar populations: very metal-poor (VMP) Galactic halo stars ($[\rm{Fe/H}]<-2.0$) from the JINAbase compilation and references therein \cite{Abohalima2018}, extra-Galactic systems \cite{Aoki2007,Ji2018}, the M5 globular cluster \cite{Lai2011}, Milky Way thin-disk \cite{delPeloso2005}, and Galactic solar twins \cite{Unterborn2015,Botelho2019}. The five metal-poor stars that have a derived abundance of uranium are from \cite{Cowan2002,Frebel2007,Siqueira-Mello2013,Placco2017,Holmbeck2018}. Zero on both axes defines the Solar system value.
 }
\label{fig:theu_observations}
\end{figure*}

The Sun and stars similar to it are young by cosmic standards.
The gas that made the Sun and its planets is the result of the preceding nine billion years of chemical evolution.
The universe started with H, He, trace amounts of Li, and negligible amounts of anything heavier.
The most massive stars---such as the first generation of stars in the universe was thought to be---continue subsequent stages of fusion in their cores up to iron.
Over time, the nearly pure H-He composition of the universe became enriched with Fe and other elements ejected by stars as they evolved.
The oldest stars, therefore, are reasonably assumed to be those with the lowest amount of Fe (low `metallicity'), since they were made at a time when the universe was still chemically primitive.

Not only can actinides be measured in the atmosphere of both our Sun and stars similar to it, but Th (and sometimes U) can also be found in stars much more chemically primitive than the Sun.
These `metal-poor' stars were made from gas that had less than 1\% of elements with $Z>=3$ than the gas from which the Sun was made.
Given that the universe evolved from a metal-free state to its current metal content, the low metallicity of these stars indicates that they are much older than the Sun, possibly even belonging to the second generation of stars ever made in the Galaxy.
Galactic chemical evolution modeling supports this basic assumption, though the correlation of metallicity with age becomes less straightforward at early times \citep[e.g.,][]{Wise2012}.

As can be seen by Fig.\ \ref{fig:generic_solar}, elements made by the \emph{r} process are a small fraction of the total composition of the Solar system.
In metal-poor stars, the abundances of heavy elements ($Z\geq 26$) are reduced even further by a factor of ten or more in comparison.
However, even though their absolute elemental abundances are low, heavy elements can still be identified in metal-poor stars \cite{Roederer2013}.
Figure \ref{fig:theu_observations} shows the abundance of Th relative to Fe (scaled to the Sun) in stars inside and outside the Galaxy spanning a range of metallicities ([Fe/H]), with the most metal-poor (lowest [Fe/H]) likely being the oldest stars.
Stars with $[\rm{Fe/H}]<-2.0$ have less than 1\% of the iron that the Sun has and are called `very metal-poor (VMP).'
The `[X/Y]' notation indicates the logarithm of the number abundance ratio of element X per Y, then scaled to the Solar ratio.
For example, stars with $[\rm{Th/Fe}]=1.0$ have ten times the number of thorium atoms per iron atom than the Solar system.
The (currently only) five stars in which U---in addition to Th---is detected \cite{Cowan2002,Frebel2007,Siqueira-Mello2013,Placco2017,Holmbeck2018} are indicated in Fig.\ \ref{fig:theu_observations} along with two cases in which Th has been identified in stars outside of the Milky Way (i.e., in the dwarf spheroidal Galaxy, Ursa Minor \cite{Aoki2007}, and in the ultra-faint dwarf galaxy, Reticulum II \cite{Ji2018}).
The presence of the \emph{r} process elements, including Th, in metal-poor stars is evidence that the \emph{r} process has been active even since early times, both in this galaxy and in others.

\begin{figure*}
 \centering
 \includegraphics[width=0.85\textwidth]{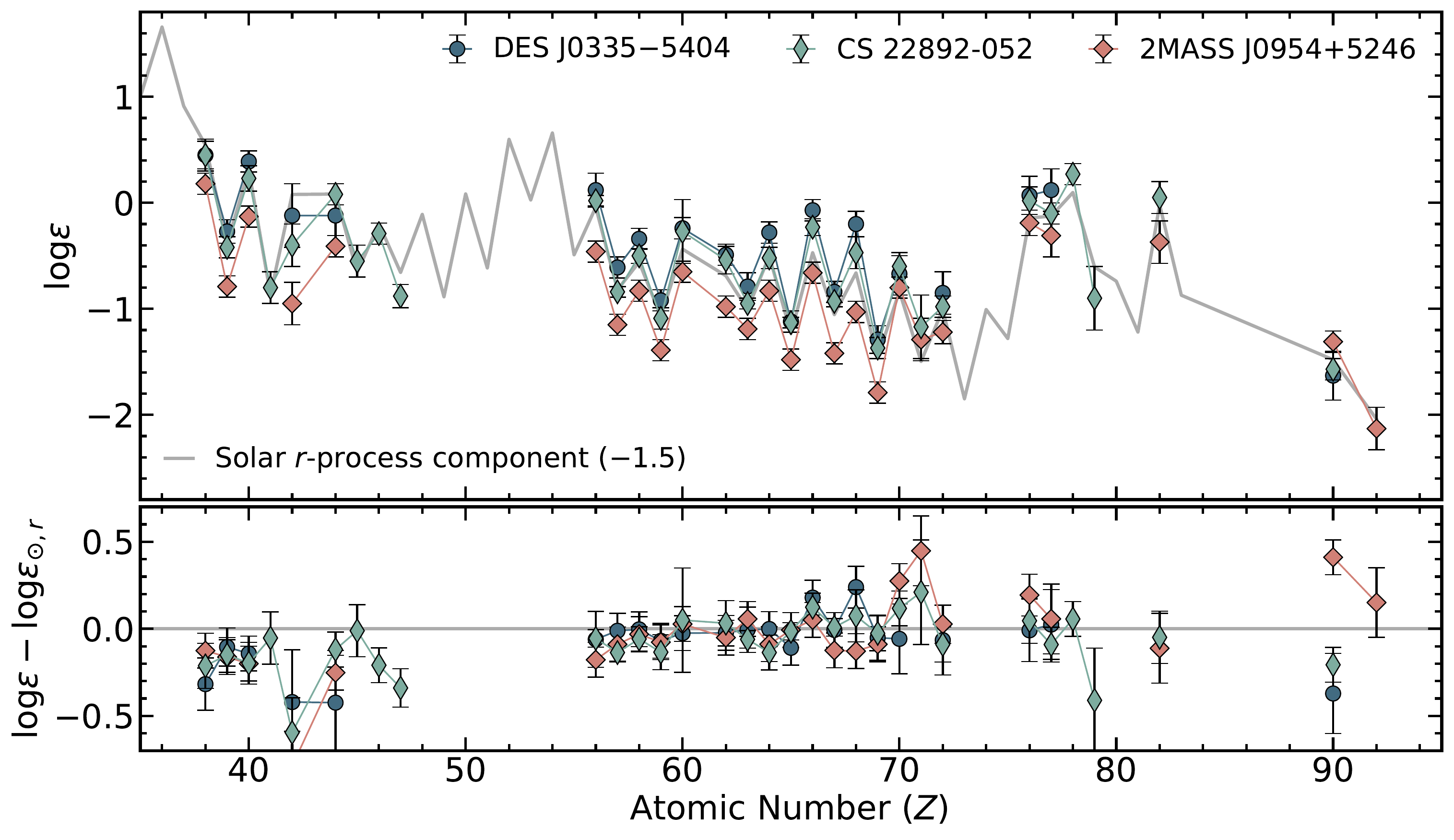}
 \caption{Top: elemental abundance patterns of three metal-poor stars with varying degrees of actinide enhancement (DES\,J0355$-$5404 \cite{Ji2018}, CS\,22892-052 \cite{Sneden2003}, and 2MASS\,J0954+5246 \cite{Holmbeck2018}) compared to the Solar \emph{r}-process component scaled by a constant. Bottom: differences between the stellar abundances and the Solar \emph{r}-process component by element, normalized to the average difference among the lanthanide elements.}
\label{fig:abundance_pattern}
\end{figure*}

The range spanned by [Th/Fe] in Fig.\ \ref{fig:theu_observations} indicates that either these elements are not produced in the same \emph{r}-process site, or not always with the same ratio every time.
If they were, the Th/Fe ratio would be constant over time (and over metallicities).
Instead, the stochasticity of thorium---especially the high abundances at low metallicities---is evidence that rare, prolific \emph{r}-process sites were active at early times in the Galaxy.
The convergence of [Th/Fe] towards the Solar value ($[\rm{Th/Fe}]=0$) by Solar metallicity ($[\rm{Th/Fe}]=0$) could be explained by the actinide-producing \emph{r}-process site becoming more frequent and less prolific (i.e., a constant production over time), or simply that the actinide abundances `average out' due to, e.g., gas mixing as the Galaxy ages.
In any case, sites that produce an abundance of actinides could possibly have produced non-trivial amounts of the superheavy elements as well.
However, due to the very old ages of stars showing the highest actinide abundances ($>$10 Gyr), any superheavy elements that were produced would have all but decayed by the time of their modern observation.
Although these stars in and of themselves offer little hope of observing the superheavy elements naturally through their starlight, their existence raises questions about the \emph{r} process site: are prolific producers of the actinide elements still active in the Galaxy today? If so, then is there a change of catching superheavy-element production in the act?

Many other \emph{r}-process elements in stars can be measured through stellar spectroscopy.
As long as a strong atomic transition occurs within visible wavelengths, it is possible to identify the corresponding absorption feature with ground-based telescopes.
For more energetic transitions with wavelengths in the ultraviolet (UV) range, space telescopes with UV capabilities must be used.
Fortunately, about half ($\sim$30) of the \emph{r}-process elements have atomic transitions in the visible range which are strong enough to allow ground-based telescopes to obtain nearly complete elemental abundance patterns of stars, similar to Fig.\ \ref{fig:generic_solar}.
Another dozen or so \emph{r}-process elements have strong transitions mainly in the UV regime, requiring space-based telescopes for even more complete abundance patterns.
Recently, 42 \emph{r}-process elements were measured in the spectrum of a metal-poor star by combining space and ground-based observational facilities, representing the most complete abundance pattern for a star outside the Solar system \cite{Roederer2022}.
Figure \ref{fig:abundance_pattern} shows three metal-poor stars in which their elemental abundances appear in an \emph{r}-process pattern, similar to the \emph{r}-process fraction derived for the Solar system.
The Solar \emph{r}-process pattern, scaled down to roughly the level of abundances in the other stars, is shown for comparison.
Note again here that their abundance patterns are necessarily elemental since most atomic absorption features are degenerate to the isotopic number of the element.
In addition, there are no measured elements with $Z=43$ (Tc), $Z=61$ (Pm), and $82<Z<90$ (everything between Pb and Th) since all known isotopes of these elements are radioactive over stellar lifetimes.
Elements with $47\lesssim Z<56$, which includes most of the second \emph{r}-process peak, and around $Z=80$ do not have strong atomic transitions within the visible range, making their detection in stellar spectra unfeasible with ground-based telescopes.
Some elements ($46\lesssim Z\leq 52$) can be obtained with UV-capable facilities (such as the Hubble Space Telescope), which has been accomplished for a handful of metal-poor stars \cite{Cowan2002,Cowan2005,Roederer2012,Roederer2012b,Siqueira-Mello2013,Peterson2020}.

Despite the tendency of elemental \emph{r}-process abundances observed in stars to show a considerable similarities---appearing to suggest a sort of universality to their production mechanism---they show significant deviations in their actinide abundance compared to the rest of their \emph{r}-process elements.
One such deviation is the small subset of stars that demonstrates a strong enhancement in their actinide abundance.
These stars have been termed \textit{actinide-boost} stars due to their enhanced levels of Th relative to the lanthanide elements (e.g., Eu), which otherwise do display remarkable universality \cite{Lai2008,Mashonkina2014,Eichler2019,Holmbeck2019a}.
Of the VMP stars only a couple dozen have thorium abundances derived, and about one-third of those display the actinide-boost signature \cite{Hill2017}.
Abundance variations like the actinide boost challenge the notion of `\emph{r}-process universality,' and some studies show that star-to-star actinide variations may be attributed to different astrophysical conditions within a single astrophysical \emph{r}-process site \cite{Holmbeck2019b}.
One such actinide-boost star---2MASS\,J0954+5246, the most enhanced yet known \cite{Holmbeck2018}---is shown in Fig.\ \ref{fig:abundance_pattern}.
Compare its Th abundance to a star considered `actinide normal' (approximately agreeing with the Solar value after accounting for its age), CS\,22892-052 \cite{Sneden2003}, and a star that is potentially actinide \emph{deficient} (much lower than Solar), DES\,J0335$-$5404 \cite{Ji2018}, belonging to the extra-Galactic dwarf galaxy Reticiulum II.

Besides the important implications for understanding the production of actinides throughout the universe, this overproduction of actinide phenomenon may, in principle, be used to explore the possibility of superheavy elements, as those astrophysical conditions producing the most actinide material could suggest the possibility that superheavy progenitors along decay series may be maximally produced as well.
Alternatively, it may be that the production of material of superheavy elements could \textit{inhibit} the production of actinides.
If nuclei along superheavy decay series exhibit strong spontaneous fission branchings, material that would otherwise be destined for long-lived actinides would be effectively rerouted into much lighter elements.
In this scenario, then, the overproduction of the long-lived actinides may be correlated with the production of \textit{fewer} superheavy elements.
In either case, advances in nuclear theory and precision measurements will both be crucial for resolving the decay modes and branching ratios that are critical to fully understand this relationship.

If superheavy elements were ever measured in stars, they would have to have half-lives comparably long to the age of the star in order to be detected through stellar spectroscopy, as they would otherwise decay on observational timescales.
Elements occupying the Island of Stability provide the most optimistic detection prospects among the superheavies, as they are projected to be the longest lived.
Unfortunately, current estimates of these half-lives (seconds to years) \cite{Karpov2012} are still far too short to support their observation over the relevant astrophysical timescales.
At the present, the only remaining feasible possibility is to attempt to catch element formation `in the act.' 
This would involve unearthing unique features in atomic or nuclear spectra for a nearby event \cite{Kasen2013,Korobkin2020,Misch2021}.
For example, a sufficiently strong $\gamma$-ray signal as part of a larger set of multi-messenger observations (e.g., gravitational waves), could distinguish between light and heavy \emph{r}-process components \cite{Timmes2019}.
A complementary approach may also be possible via indirect observation that provides a distinct signature of the influence of long-lived decay products of the heaviest elements \cite{Wanajo2014,Zhu2018,Wu2019,Zhu2021,Barnes2021,Terada2022}.

\subsection{Fission fragments}

Calculations that produce superheavy elements in an \emph{r} process typically find that the longest timescales for the populated nuclei are on the order of days to years \cite{Petermann2012,Panov2018}.
As discussed above and in Sec.\ \ref{sec:nucphys}, these short half-lives and low fission barriers in the transactinide region reduces the chances of directly detecting the superheavy elements and/or their lower-charge progenitor nuclei.
However, fission could be harnessed as a way to infer the existence or production of the superheavy elements.
The fission products of heavy nuclei populate lower mass and charge numbers, and---after finding their own way back to stability via $\beta$ decay---eventually lead to an enhanced production of lighter nuclei (around the second \emph{r}-process peak $A\sim 130$).
Since all but two $Z\leq 84$ elements have at least one stable isotope, such an overproduction from fission deposition can persist on indefinite (i.e., observable) timescales.
Substantial overproduction of key elements in the fission product region could therefore be observed in, e.g., stars and thus would indicate that fission of heavy or superheavy elements must have occurred.

The particular elements that would unequivocally be evidence of fission by (super)heavy nuclei are uncertain.
The landscape of fission barrier heights determines where the bulk of heavy nuclei undergo fission, and this region is highly uncertain and unconstrained, as discussed in Sec.\ \ref{sec:fission} \cite{Vassh2019,Giuliani2020}.
Similarly, the fragment distribution of the extremely neutron-rich nuclei undergoing fission in the \emph{r} process is also unknown \cite{Mumpower2020}.
Theoretical descriptions of fission fragment deposition vary from producing lower-mass nuclei around $A\sim 105$ to being responsible for the rare-earth peak at $A\sim 165$ \cite{Goriely2013,Shibagaki2015,Vassh2021}.
In any case, if uncertainties on the fission rates (from the fission barrier heights) and fission fragment distributions can be reduced, key elements observed in metal-poor stars can be associated with and even quantify the extent of overproduction via fission e.g., through network tracing \cite{Sprouse2021}.

\subsection{Beyond uranium}

\begin{figure*}
 \centering
 \includegraphics[width=0.9\textwidth]{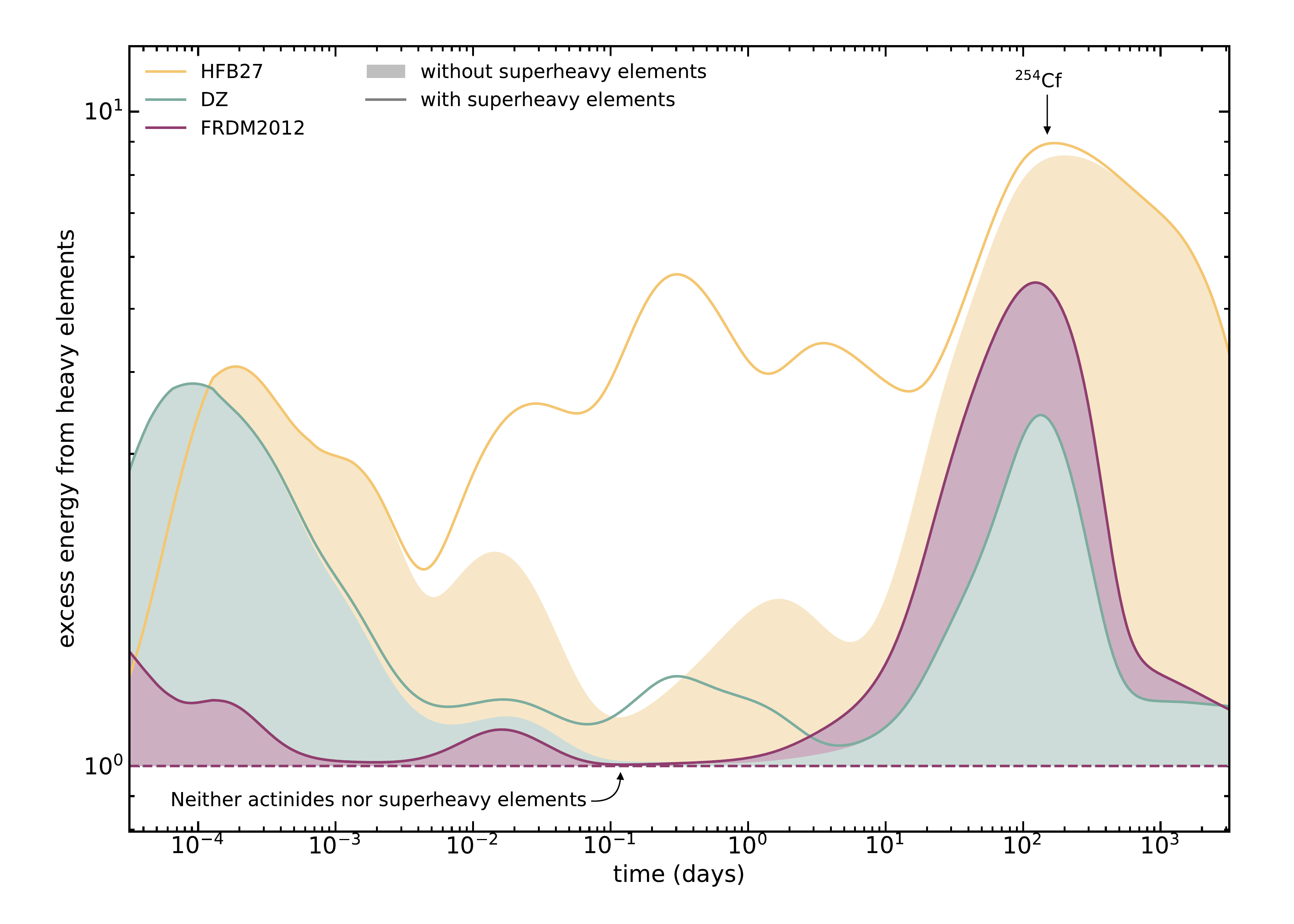}
 \caption{Enhancement to energy released from the $\alpha$ decay and fission of very heavy and superheavy elements. Standard \emph{r} process conditions (cold, very neuron-rich) were run using three nuclear mass models and fission barrier heights: HFB with HFB-14 barrier heights \cite{Arnould2006} (yellow), DZ \cite{Duflo1995} with KTUY \cite{Koura2014} (green) and FRDM2012 \cite{Moller1995,Moller2012,Moller2016} with FRLDM \cite{Moller2015} (pink). The late-time bump around 130 days that appears in all three cases is attributed to the spontaneous fission of $^{254}$Cf.}
\label{fig:heating_rates}
\end{figure*}

Although the presence of actinide elements are an indication that the \emph{r} process can occur naturally, their existence alone is insufficient to determine the full elemental extent of \emph{r}-process production.
The most promising avenue towards potentially confirming the production of superheavy elements in nature is through direct observation of \emph{r}-process events, especially in the light that is emitted by radioactive nuclei produced therein.
High-opacity lanthanides and actinides affect the bulk properties of `kilonova' light (see \cite{Metzger2019} for a recent review) or by producing an observable feature in their spectra \cite{Watson2019}.
Following kilonova and gamma-ray bursts on short timescales after an \emph{r}-process producing event could reveal fingerprints of extremely radioactive nuclei with short half-lives.
For example, it has been found that the radioactive decay of $^{254}$Cf ($t_{1/2}\approx 60$ days) can leave a distinguishable feature in the light curve following a nucleosynthetic event \cite{Zhu2018}.
The energy released by the (spontaneous) fission of $^{254}$Cf was initially thought to be responsible for powering the light curves of Type Ia supernovae \cite{Burbidge1956}.
However, as supernovae have fallen out of favor as actinide-producing \emph{r}-process sites, $^{254}$Cf has resurfaced as theoretically contributing to kilonova light curves following, e.g., the merging of a compact-object binary system \cite{Wanajo2014,Zhu2018}.
Zhu \textit{et al.} found that the fission of $^{254}$Cf produces sufficient energy to delay the otherwise power-law decay of kilonova lightcurves at about 60--100 days post-merger.

Figure \ref{fig:heating_rates} shows the impact of $\alpha$ decay and fission of the heavy elements on the heating rate from material produced by the \emph{r} process.
This `heat' tracks the energy produced by nuclear reactions and decays (see Sec.\ \ref{sec:nucphys}) and, through the opacities of atoms constituting the material, bears directly on the observable lightcurve from such an event.
A standard \emph{r}-process simulation is used with three nuclear physics combinations that optimistically and pessimistically produce superheavy elements, according to Fig.\ \ref{fig:Zchains}: DZ, HFB27, and FRDM2012.
As a baseline case, the total energy release as a function of time is taken and any contribution by $\alpha$ decay and fission from elements with $Z\geq 89$ is removed, which includes all the actinides and superheavy elements.
The total energy release is then recalculated, this time including the contribution from all actinides, but no superheavy elements (e.g., only $Z\leq 103$).
Lastly, all elements are allowed to contribute to the energy release.

To visualize the impact from all $Z\geq 89$ and $Z\geq 104$ elements, we compare the two latter calculations that have $Z\geq 89$ contribution to the first calculation that does not include these elements.
In Fig.\ \ref{fig:heating_rates}, the shaded regions show the effect that simply including actinides (but no superheavy elements) has on the total energy release.
The solid line shows the effect when superheavy elements (in addition to the actinides) are included.
When any of the $Z\geq 89$ elements are included, the energy released is increased relative to the case without those elements, as a consequence of the energetic $\alpha$ decay and fission of those elements now being allowed to add energy (heat) to the \emph{r}-process ejecta.
Note that the three mass models qualitatively agree on when the most significant contribution occurs, but disagree in their details of relative impact.

The most interesting comparison is between the solid lines and filled regions in Fig.\ \ref{fig:heating_rates}.
For the most part, the solid and filled curves are indistinguishable (especially for FRDM2012), meaning that superheavy elements are not contributing significantly more than just the actinides themselves.
A revisit of Fig.\ \ref{fig:Zchains} explains this result; the FRDM2012 mass model most pessimistically produces the superheavy elements, so their low abundances have little to no impact on the total energy release.
On the other hand, HFB27 produces the highest total abundance of superheavy elements at that time, though that abundance is concentrated at $Z=104$ (Rf).
Notably, the excess energy provided by the production of superheavies is on the order of a factor of two.
The DZ model shows a lower total superheavy abundance than HFB27, but a greater extent in charge number, peaking at $Z=108$ (Hs).
Consequently, HFB27 and DZ do show a deviation between the cases with and without superheavy elements (see around 0.1 and 1 day in Fig.\ \ref{fig:heating_rates}), presenting a chance to witness their natural, astrophysical production.
These energy-excess features---whether from solely the actinides or the superheavy elements---would reveal themselves in both the bolometric light curves of \emph{r}-process events and the near-infrared ($JHK$) bands \cite{Wu2019}.
Observations of late-time infrared excess in lightcurves could therefore indicate the presence of actinides and superheavy elements \cite{Zhu2021}.
Such observations may be possible with the NIRSpec detector on the recently launched JWST, which has observational capabilities in the near-infrared region \cite{Ferruit2022}.
Long-term follow up of future kilonova events can be a route to observationally confirming the extent of heavy-element production by the \emph{r} process.

Not only would such observations directly confirm the synthesis of actinide and possibly superheavy material by the astrophysical \emph{r} process, but they also have the potential to reveal unknowns of the \emph{r} process itself.
Recent studies have explored the impact that nuclear physics far from stability has on the production of $^{254}$Cf and other short-lived nuclei that can contribute to radioactively powered light curves \cite{Vassh2019,Barnes2021,Zhu2021}.
For example, the choice of theoretical basis for calculating unknown nuclear masses and reaction rates determines how much material is lost to fission before the actinides, superheavies, and their $\beta$-decay progenitors are populated by the \emph{r} process.
On Fig.\ \ref{fig:heating_rates}, the contribution to the energy excess from the spontaneous fission of $^{254}$Cf can be seen in all models at around 130 days.
Depending on the reality of nuclear physics far from stability, fission of short-lived superheavies could leave a similar signature on the electromagnetic counterparts to \emph{r}-process events, albeit on shorter timescales than in the case of $^{254}$Cf, see, e.g., the superheavy-element features in the HFB and DZ models in Fig.\ \ref{fig:heating_rates} at $t\lesssim 1$~day.
Even if features attributable to actinides \emph{only} and not superheavies are identified, such non-detections could provide necessary constraints on nuclear properties far from stability.

There is also the possibility of observing actinides (or superheavy elements) in spectra.
An absorption feature of $_{38}$Sr was likely detected \cite{Watson2019} in the light following the historic 2017 detection of a merger of two neutron stars \cite{Abbott2017a,Abbott2017b,Coulter2017}.
Although Sr is a much lighter element than the actinides or superheavies, there nonetheless exists the possibility that absorption features due to the heaviest elements (or their fission fragments) could be directly detectable in similar ways.
In lieu of a detailed spectra and light curves at late times, other signature of fission can be found, and therefore signatures of translead and transactinide production.
The fission of heavy elements can release high-energy (a few MeV) gamma rays, distinguishable from competing gamma rays released by the $\beta$-decay of radioactive nuclei in the same material \cite{Wang2020}.
These prompt gamma rays could be observed from Galactic \emph{r} process events by next-generation MeV detectors (e.g., AMEGO, e-ASTROGAM, or LOX \cite{Timmes2019}).

\section{Summary \& future outlook}\label{sec:summary}

In this review we have outlined the synthesis of the heaviest elements in the rapid neutron capture (\emph{r}) process from the perspective of nuclear physics and astrophysical observations.
We have covered the properties of heavy and superheavy nuclei, the reactions and decays that are involved in creating and destroying those elements, the astrophysical conditions which favor their production, and observations of the heaviest elements.
Although seemingly disparate on the surface, a unified understanding of theory, experimental physics, and observational astronomy is necessary to answer longstanding questions regarding the production of the heaviest elements in the \emph{r} process.%

Under the most favorable conditions that allow synthesis of the heaviest elements, the \emph{r} process can undoubtedly create nuclei that possess copious numbers of neutrons and/or protons.
None of these nuclei have measured properties, resulting in a reliance on nuclear theory when simulating nucleosynthesis.
Current theoretical models exhibit a wide variation in predictions, ranging from no superheavy element creation to modest abundance on the order of actinide production.
For superheavy elements to be synthesized in the \emph{r} process, a combination of suitably neutron-rich astrophysical conditions must be paired with suppressed fission likelihoods for a large swath of the heaviest nuclei.

To further make progress regarding the synthesis of the heaviest elements in the \emph{r} process, it will be necessary to compile and incorporate the latest nuclear data into simulations \cite{Kolos2022}.
The propagation of uncertainty in nuclear properties \cite{Martin2016,Sprouse2020} as well as sensitivity studies \cite{Aprahamian2014,Mumpower2014,Surman2014} are valuable tools for the community as they motivate experimental measurements and provide additional insight into the \emph{r} process \cite{Vassh2022}.
Future measurement campaigns (e.g., those enabled by current generation fragmentation facilities \cite{Vilen2018,Tain2018,Wu2020}) will be valuable in determining the properties of lighter nuclei (which may be fission products of heavier species).

The study of nuclear properties in the actinide region requires alternative techniques from fragmentation.
Multi-nucleon transfer reactions may provide an efficient method to produce and study neutron-rich nuclei in this region \cite{Zagrebaev2008,Loveland2019}.
Some models of multi-nucleon transfer predict large cross sections to produce the actinides with diminishing cross sections for the transactinides \cite{Yanez2015}.
Production of isotopes not far from what has been measured may provide valuable information for nucleosynthesis studies of long-lived, potentially observable actinides, e.g., the propensity of neutron-rich actinides to undergo $\beta$-delayed fission \cite{Mumpower2022}.
More generally, the use of heavy beams (e.g., of $^{238}$U) can help to elucidate the unknown physics of the heaviest elements, such as their masses, half-lives, or fission barriers.
Facilities like the recently commissioned `$N=126$ factory' at Argonne National Laboratory offer unique capabilities for the production of these exotic neutron-rich nuclei \cite{Savard2020,Valverde2020}.

The study of the production of the heaviest elements in the \emph{r} process intimately ties in with astrophysical observations, and the next generation of observing campaigns will continue to strengthen this bond.
Future multi-messenger observations (e.g., $\gamma$-ray \cite{Timmes2019} and other space-based missions like LISA \cite{AmaroSeoane2022}) will provide critical new abilities to observe and interpret aspects of the fundamentally multi-physics problem of heavy-element creation.
As gravitational wave detection capabilities grow, an increasing demand will be placed on electromagnetic transient identification and followup.
Capabilities of next-generation telescopes will be fundamental for kilonova followup and may even be able to catch superheavy-element production in action, provided that the excess in heating rate at 0.1--1 day leaves a sufficiently distinguishable signature on transient light curves.
At the same time, recent advances in the theoretical prediction of actinide and superheavy opacities may offer a path forward for resolving superheavy vs.\ actinide production in electromagnetic observations of kilonova environments \cite{Fontes2022}.
If the Island of Stability is populated in \emph{r}-process sites, their comparatively long half-lives (up to years) may broaden this observational window and allow detections on even longer timescales.
A robust study on the detectability of superheavy elements in kilonova light curves would be a worthwhile exploration.

As we move forward with both multi-messenger and kilonova-followup observations, it will be especially important to understand the MeV background from our vantage point in the universe \cite{Bloser2022}, as these high-energy signals may provide an important \textit{in situ} signature of nascent synthesis of the heaviest nuclei \cite{Korobkin2020,Wang2021}.
However, we need not look only to future astrophysical events that produce these nuclei; recent and continuing observations of the oldest stars can provide pristine snapshots of heavy element production associated with \emph{r}-process events \cite{Hansen2018,Sakari2018}.

Whether superheavy elements can be made naturally in the cosmos and how we can observe them---if at all---is one of the critical questions driving research in this exciting field of study.
Continued work across these domains of physics will ultimately sustain a tightly woven network of research efforts \cite{Schatz2022}.
Taken as a whole, these efforts will help us uncover the mysteries of the heaviest elements in nature.

\begin{acknowledgements}
M.R.M.\ gratefully acknowledges the stimulating discussions that ensued at the 4th International Symposium on Superheavy Elements (SHE2019) for inspiring this work.
Conversations with Filip Kondev, Hiroyuki Koura, Gabriel Martinez-Pinedo, Witek Nazarewicz, Yuri Oganessian, and Jorgen Randrup were particularly inspiring.
Support for this work was provided by NASA through the NASA Hubble Fellowship grant HST-HF2-51481.001 (E.M.H.) awarded by the Space Telescope Science Institute, which is operated by the Association of Universities for Research in Astronomy, Inc., for NASA, under contract NAS5-26555.
T.M.S and M.R.M were supported by the US Department of Energy through the Los Alamos National Laboratory (LANL). LANL is operated by Triad National Security, LLC, for the National Nuclear Security Administration of U.S.\ Department of Energy (Contract No.\ 89233218CNA000001).
\end{acknowledgements}

\bibliographystyle{spphys}    %
\bibliography{refs.bib}  %

\end{document}